# Experimental identification of multimode traveling waves in a coupled wave-tube

Yoav Vered[a] and Izhak Bucher

Dynamics Laboratory, Faculty of Mechanical Engineering, Technion – Israel Institute of Technology,

Haifa 3200003, Israel

## Abstract

An essential part of nondestructive testing and experimental modeling of waveguides is the decomposition of propagating wave patterns. The traveling wave ratio is a measure of partial reflections assisting in quantifying the pureness of a single traveling wave from a power flow perspective. This paper expands the notion of traveling wave ratio for multimode systems and outlines several schemes capable of decomposing the waves into their different traveling modes while quantifying their traveling and standing proportions individually. A method to strike an optimal balance between increasing model order and to maintain low uncertainty is proposed. An experimental study performed on an acoustic wave tube, which utilizes the various methods while assessing their accuracy and performance, is reported. The results described here emphasize the importance of including additional propagating modes. In addition, the results illustrate the capability of using the recursive multichannel least-mean-squares method for both a fast decomposition and as a basis to formulate closed-loop schemes controlling the wave's propagation patterns.

---

[a] Electronic mail: syoavv@gmail.com ; syoavv@campus.technion.ac.il

# Graphical abstract

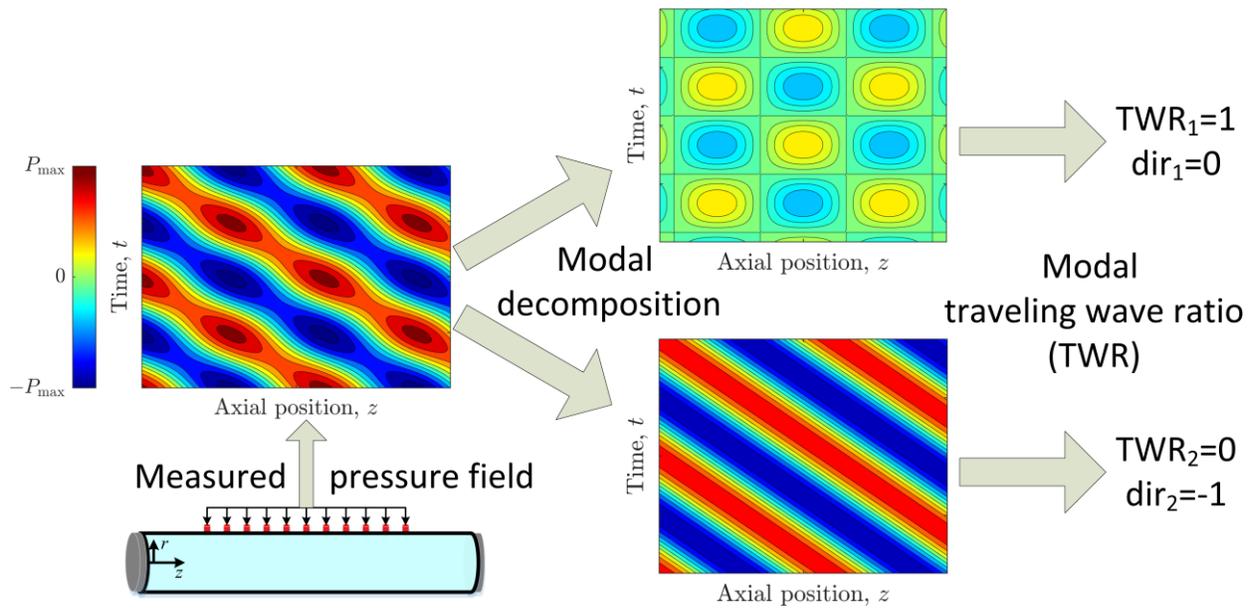

# 1. Introduction

The analysis of continuum dynamics using a basis of traveling waves has been employed in various fields. Few notable examples are Acoustic, Elastic, Electromagnetic, Optics, and Quantum mechanics. While the simple models of propagation give rise to a single propagating mode, a realistic model results in several simultaneous propagating modes. Thus, multiple modes can exist at the entire frequency range [1,2] or, in some cases, only above the so-called cutoff frequency [3].

Acoustic wave tubes such as the impedance tube, standing wave tube, and pulse tube, play an important role in the conduction of nondestructive testing (NDT) of acoustic properties [4,5]. The wave tube is composed of a hollow elastic cylinder and is filled with the required fluid. Normally, as part of the analysis, the propagating pressure wave is decomposed into the forward and backward traveling wave under the assumptions of a single and planar propagating mode [6]. In reality, an acoustic wave tube that contains a compressible fluid surrounded by an elastic enclosure gives rise to dispersion relation describing multiple propagation modes (referred to as multimode propagation [1]) at all frequencies [1,7]. This is true for air-filled wave tubes [8] and liquid-filled wave tubes [2,9], where the coupling is stronger. In addition, when considering air-filled tubes, it is common to neglect the propagation attenuation factor since it was found to be much smaller than the boundaries' attenuation [8,10]. The latter is no longer necessarily true for liquid-fluid wave tubes where nonnegligible propagation attenuation usually takes place. Thus, resulting in complex modal wavenumbers [9,11,12] for the propagating modes.

Each of the propagating mode branches of the dispersion relation is the outcome of a different energy conveying mechanism [13], whose contribution to the total transfer of energy can be quantified by using the modal power ratio [14]. The modal power ratio for fluid-filled acoustic waveguides was previously employed to describe the pressure distribution along straight cylindrical acoustic waveguides [14], curved acoustic waveguides [15], and straight axisymmetric general cross-section acoustic waveguides [16].

The standing wave ratio (SWR) [17] and traveling wave ratio (TWR) [18,19] have been used to quantify the direction and purity of the propagating mode for the single-mode scenarios and to identify the acoustic impedance in a standing wave tube [20]. The power ratio and the TWR are closely related [21]; thus, either one can be used to quantify the fraction of propagating energy in each conveying mechanism. In a wave control application, TWR is preferred over SWR because the former is always finite and thus more suitable for real-time control [18].

Traveling wave control schemes [22] are mostly developed based on a single propagating mode in each propagation direction. The existing control schemes were developed and applied for many elastic and acoustic waveguides such as the cases of 1D beams [23–25], lumped flexible systems [26–28], systems governed by the 1D wave equation [18,28,29], and systems governed by the 2D wave equation[30–32]. As discussed in [18], the open-looped model is highly sensitive to model uncertainties; therefore, the traveling wave's experimental identification approach is crucial for reasonable control. The decomposition of a single traveling wave into its forward and backward components was done previously, both in a batch mode [18,24,33] and recursively [17,21], and is now a standard analysis tool. A multimode full wave decomposition was suggested in [34,35] when a rigid boundary tube was considered.

In a previous work [36], the generalization of the TWR to the multimode scenario was used to control the TWR of the principal acoustic mode [1] by using a batch-based identification procedure. The present work develops a generalization of the TWR using a projection of the pressure wave onto the modal wave function basis and outlines tools enabling recursive and fast identification of each modal TWR. Four methods to identify the TWR in the multimode scenario are outlined: the first, a batch multichannel least-squares (MC-LS) method; the second, a recursive multichannel recursive-least-squares (MC-RLS) [37] method; the third a recursive multichannel least-mean-squares (MC-LMS) method [37]; the fourth, an extended recursive synchronous demodulation method [38]. While the first three methods are model-based and rely on the knowledge of the dispersion relation of all propagating modes, the fourth method relies on the dispersion relation of a single-mode but is limited by the number and distribution of sensors. The dispersion relation can be modeled analytically [2,7,9], numerically [39,40], or experimentally [8].

Least-squares (LS) is a conventional method and tool for the model-based identification technique [41]. An extensive introduction to the method and its generalization is found in [42]. A thorough error analysis of the LS method using parameter perturbations was given by Björck [43]. The recursive formulation of the LS problem and other model-based methods for active control are also common knowledge nowadays. A fundamental reference on recursive adaptive estimation and control is [44], serving as a guideline to the present work. The number of propagating modes is not known a priori; therefore, there is a need to avoid overfitting [42] to minimize errors. Overfitting can be avoided by employing an L-curved analysis [45,46] to choose the correct model-order (number of propagating modes).

Described in this work is a numerical and experimental study of state-of-the-art modeling techniques for recursive modal decomposition in an acoustic wave tube. The paper is organized as follows: Section 2 presents and discusses the fundamental work of Del Grosso [1] to introduce the concept of multimode propagation. Section 3.1 introduces the TWR based on the rigid wall assumption. Section 3.2 generalizes the TWR by projecting the pressure wave onto the modal wave basis based on the elastic acoustic wave tube model. Section 4 outlines methods to experimentally identify the modal TWR (MC-LS, MC-RLS, MC-LMS, synchronous demodulation), subsection 4.1.1 and 4.1.2 includes a discussion concerning the possible error sources in the model inversion and outlines requirements on sensor positioning. Section 5 presents a sensitivity-based analysis chart to avoid overfitting and choose the model order. Section 6 presents a numerical comparison between the methods. Section 7 presents the experimentally obtained results, subsection 7.1 describes the experimental air-filled acoustic wave tube, subsection 7.2 presents the experimentally sensitivity-based analysis, subsection 7.3 depicts the method comparison using experimentally obtained data, and subsection 7.4 presents a comparison between the batch MC-LS and the recursive MC-LMS based on the modal TWR contour maps. The main conclusions of this paper are summarized in Section 8.

## 2. Motivation – multimode propagation in an acoustic wave tube

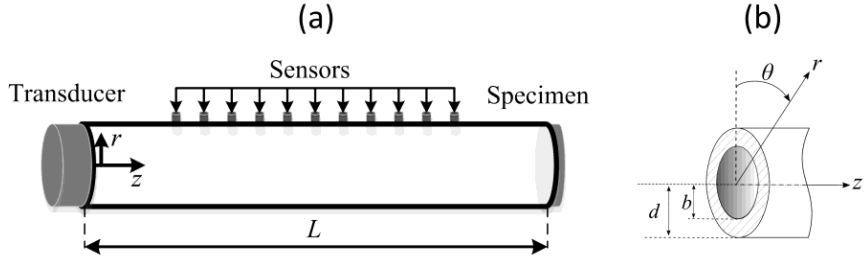

Fig. 1 – (a) Model of a coupled acoustic-elastic wave tube similar to the one used for NDT of acoustical properties. (b) A cross-section model of a coupled acoustic-elastic waveguide used for dispersion analysis.

Acoustic wave tube, illustrated in Fig. 1a, is a common experimental system to conduct NDT of acoustic properties for different materials and environments; a comprehensive review about their use and method of operation can be found in [6]. The use of a representative physical model composed of forward and backward traveling waves is the heart of the different evaluation methods. Most methods neglect the acoustic-elastic coupling by employing a rigid boundary condition model for the elastic tube. The rigid boundary model results in a single nondispersive planar propagating mode up to the first cut-off frequency [3]. On the other hand, models that incorporate the acoustic-elastic coupling give rise to a rich, multimode dispersion relation, as is shown here for the case of a 3D axisymmetric model. The governing equations for the axisymmetric acoustic-elastic waveguide model at any cross-section are [1,47]

$$\left(\frac{1}{c_1^2}\frac{\partial^2}{\partial t^2} - \nabla^2\right)\Phi(r,\theta,z,t) = 0, \ r \in (0,b), \tag{1}$$

$$\left(\frac{1}{c_c^2}\frac{\partial^2}{\partial t^2} - \nabla^2\right)\vartheta(r,\theta,z,t) = 0, \ r \in (b,d), \tag{2}$$

$$\left(\frac{1}{c_s^2}\frac{\partial^2}{\partial t^2} - \nabla^2 + \frac{1}{r^2}\right)\psi_\theta(r,\theta,z,t) = 0, \ r \in (b,d), \tag{3}$$

where $b$ and $d$ denote the inner and outer radius of the waveguide (tube), $\Phi$ is the acoustic velocity potential function [1], and $\vartheta$ and $\psi_\theta$ are the scalar potential and azimuthal element of the vector potential of the elastic displacement field used in the Helmholtz decomposition $\mathbf{u} = \nabla\vartheta + \nabla \times (0, \psi_\theta, 0)$ [47]. $c_1$ denotes the intrinsic speed of sound in the fluid, $c_c = \sqrt{(2\mu + \lambda)/\rho_s}$ and $c_s = \sqrt{\mu/\rho_s}$ denote the velocities of longitudinal and transverse waves in the solid. $\rho_s$ denotes the solid density, and $\lambda$ and $\mu$ are the Lamé parameters of the solid medium. The axially symmetric potential solutions of Eq. (1)–(3) are [1, Eq. (23)]

$$\Phi = \phi_{0m} J_0\left(\sqrt{k_1^2 - k_{0m}^2}\, r\right) \exp\left(i(\omega t - k_{0m} z)\right), \tag{4}$$

$$\vartheta = \left(A_{0m} J_0\left(\sqrt{k_c^2 - k_{0m}^2}\, r\right) + B_{0m} Y_0\left(\sqrt{k_c^2 - k_{0m}^2}\, r\right)\right) \exp\left(i(\omega t - k_{0m} z)\right), \tag{5}$$

$$\psi_\theta = \left(C_{0m} J_1\left(\sqrt{k_s^2 - k_{0m}^2}\, r\right) + D_{0m} Y_1\left(\sqrt{k_s^2 - k_{0m}^2}\, r\right)\right) \exp\left(i(\omega t - k_{0m} z)\right), \tag{6}$$

for angular frequency $\omega$, modal wavenumber $k_{0m}$, mediums wavenumbers $k_1$, $k_c$, and $k_s$ which are defined by the ratio $\omega/c_\circ$, where $\circ$ stands for either 1, $c$, or $s$ accordingly. Also, $J_n$ and $Y_n$ are the $n^{th}$ order Bessel function of the first kind and second kind. The constants $\phi_{0m}$, $A_{0m}$, $B_{0m}$, $C_{0m}$, and $D_{0m}$ are obtained from the boundary conditions: fluid pressure equals normal radial stress at the inner radii of the solid; fluid radial normal velocity equals that of the solid at the inner radii; zero solid radial shear stress at inner radii; zero solid radial shear stress at outer radii; zero solid radial normal stress at the outer radii. Note that the $+i\omega t$ convention is used contrary to Del-Grosso [1], use of $-i\omega t$, leading to the change in the exponential sign. By substituting the axially symmetric potential of Eq. (4)–(6), a homogenous linear system of equations can be written for the 5 constants, taking the matrix determinant yields the frequency-dependent dispersion equation in the modal wavenumber $k_{0m}$ [2, Eq.(5)]. The characteristic equation's real zeros are the propagating modes' axial wavenumbers (their complex counterparts are the evanescent modes' wavenumbers). Fig. 2 presents the analytically computed dispersion curve of an air-filled wave tube used as part of this research based on the parameters provided in Table 1.

Table 1– List of properties and their values of the wave tube used for the analytical derivation and the experimental modal decomposition.

| Property | Symbol | Value, units |
|---|---|---|
| Air density | $\rho_1$ | 1.1839, kg m$^{-3}$ |
| Air intrinsic speed of sound | $c_1$ | 346.13, m s$^{-1}$ |
| PMMA density | $\rho_s$ | 1220, kg m$^{-3}$ |
| PMMA Lamé's first parameter | $\lambda$ | 3557, MPa |
| PMMA Lamé's second parameter | $\mu$ | 1123, MPa |
| Tube length | $L$ | 2.011, m |
| Tube inner radii | $b$ | 0.042, m |
| Tube outer radii | $d$ | 0.045, m |

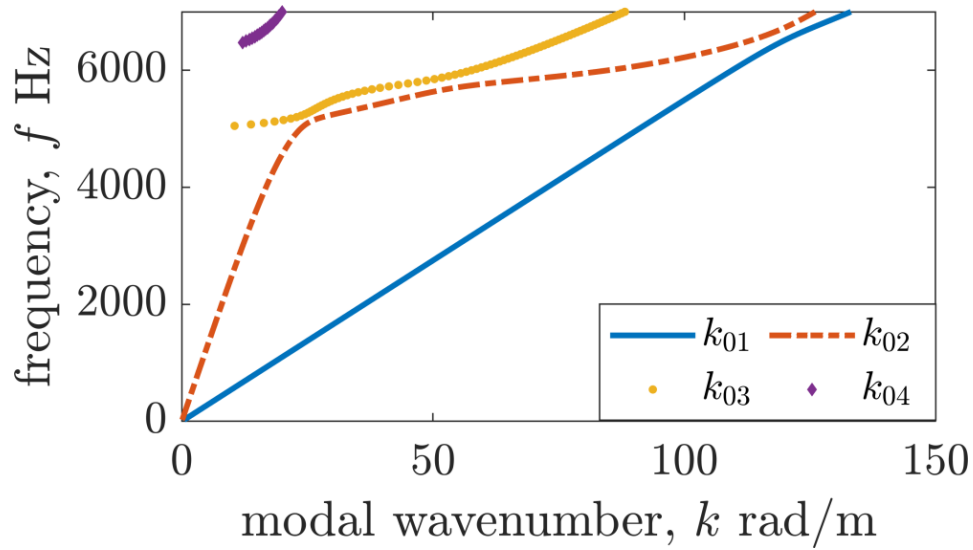

Fig. 2 – (color online) Analytical dispersion curves of an axisymmetrical acoustic waveguide.

Each branch (line) of the dispersion curve is associated with a different propagating mode and therefore acts as different energy conveying mechanism [16]. It can be seen that at all frequencies, at least two propagating modes exist; at higher frequencies, additional modes begin to propagate. Each of these frequencies is the cut–off frequency of the associated mode. Similar to the discussion in [1], it is noted that the number of real zeros of the characteristic equation is finite in the axisymmetric case. This leads to the conclusion that at every frequency, there is a finite number of propagating modes. The other, infinite number of complex zeros are associated with evanescent modes. The dispersion curves presented in Fig. 2 do not include all energy conveying mechanisms since the model neglects non-axisymmetric modes that exist in reality. A full analytical model of the 3D waveguide dispersion relation can be found in [7]. In which additional propagation modes are predicted in the low-frequency region. Still, the axisymmetric model [1] discussed here can still predict that several modes propagate at all frequencies and that the number of propagating modes is finite.

In a previous paper [8], the experimental identification of the dispersion curves of the wave tube in use here was conducted using the two-actuator phase-perturbations (TAPP) method resulting in a strong agreement with both the axisymmetric analytical model of [1] (which is presented above) and the flexural beam deflection model of a circular cross-section Timoshenko's beam [47,48]. Therefore, the need to decompose the measured pressure field into all propagating modes, which affect the accuracy from an energy perspective, is crucial to the success of acoustical NDT methods.

## 3. Traveling wave ratio

The traveling wave ratio is a scalar measure of the propagating versus circulating energies ratio [21]. It serves as a basis for quantifying these proportions on a realistic, physical model as outlined below.

### 3.1 Background – single-mode propagation

When considering the case of an axisymmetric acoustic wave tube and modeling the elastic tube as a rigid boundary, the general solution of the fluid velocity potential of Eq. (4) holds, but the boundary condition in this case becomes

$$\left.\frac{\partial \Phi}{\partial r}\right|_{r=b} = 0 \Rightarrow J_1\left(\sqrt{k_1^2 - k_{0m}^2}\,b\right) = 0 \,, \tag{7}$$

which defines the characteristic equation of the waveguide whose zeros are the modal wavenumbers, $k_{0m}$. Since $k_{0m} = k_1$ is a zero of the characteristic equation with a constant radial distribution ($J_0(0) = 1$). It follows that the principal acoustic mode, in this case, is a planar traveling wave. This is also the only real solution until the first cut-off frequency, defined by

$$f_{c,2} = \frac{\xi_{0,2}}{2\pi b} c_1, \tag{8}$$

where $\xi_{n,m}$ represents the $m^{th}$ zero of $J'_n$. Therefore, the pressure field can be expressed, up to this frequency as a one-dimensional wave by means of the D'Alembert solution [49]

$$p(z,t) = \text{Re}\{P^+ \eta^+(z,t) + P^- \eta^-(z,t)\}. \tag{9}$$

Here $P^+$ and $P^-$ denote the forward and backward propagating wave amplitude, accordingly, and $\eta^\pm$ denotes the forward and backward traveling wave functions

$$\eta^\pm(z,t) = \exp(i(\omega t \mp kz)). \tag{10}$$

The dispersion relation, in this case, is linear $\omega = c_1 k$. Which is referred to as a non-dispersive wave where both the group and phase velocities are kept constants as a function of angular frequency (wavenumber).

The Standing Wave Ratio [17] of the pressure wave is thus defined as

$$\text{SWR} \equiv \frac{|P^+| + |P^-|}{|P^+| - |P^-|}. \tag{11}$$

Since the SWR is unbounded when approaching a pure standing wave, a more usable quantification may be found in the Traveling Wave Ratio (TWR) [19]

$$\text{TWR} \equiv 1 - |\text{SWR}^{-1}| = 1 - \frac{||P^+| - |P^-||}{|P^+| + |P^-|}. \tag{12}$$

The TWR represents the ratio between the standing waves, and the total maximal amplitude of the combined traveling and standing waves, which best suits traveling wave optimization procedures as TWR=0 means a pure traveling wave while TWR=1 a pure standing one. The sign of the numerator argument defines the traveling wave's direction of propagation

$$\text{dir} \equiv sign\left(\left|P^+\right|-\left|P^-\right|\right) = \pm 1, \tag{13}$$

a positive sign means a forward traveling wave and vice-versa.

Thus, for the planar single-mode scenario, the TWR can and has been used to model the dynamic and control it [30]. In the more realistic case of multimode propagation, as presented in the motivation section, the presented definition of the TWR is not usable, and a rigorous generalization is needed.

**3.2 Traveling wave ratio generalization for a multimode scenario**

In this subsection, the generalized, multimode case of the TWR is presented. By considering the case of a multimode propagation that is predicted by the analytical solution of the elastic acoustic waveguide, several waves having different wavenumbers arise. In this case, the model of Eq. (9) is no longer adequate, and the solution needs to be decomposed into a modal summation [1,33], via

$$p(r,\theta,z,t) = \sum_{n=-\infty}^{\infty} \sum_{m=1}^{\infty} p_{nm}(r,\theta,z,t), \tag{14}$$

$$p_{nm}(r,\theta,z,t) = \text{Re}\left\{g_{nm}(r,\theta)\eta_{nm}(z,t)\right\} \quad \eta_{nm}(z,t) \equiv P_{nm}^+ \eta_{nm}^+(z,t) + P_{nm}^- \eta_{nm}^-(z,t). \tag{15}$$

Here $p_{nm}$ represents the propagation mode, $\eta_{nm}$ the modal wave, $n$ and $m$ denotes the azimuthal and radial mode order respectively, $z$ denotes the wave propagation direction as before, and $r$ and $\theta$ are the cross-section polar coordinates. $g_{nm}$ stands for the cross-section amplitude modulation function. For circular waveguide

$$g_{nm}(r,\theta) = \frac{1}{E_{nm}} J_n\left(\sqrt{k_d^2 - k_{nm}^2}\, r\right) e^{in\theta}, \tag{16}$$

where $k_d$ is the associated medium wavenumber of each potential function as defined in the motivation section, and $E_{nm} = \left(2\pi \int_0^b J_n\left(\sqrt{k_d^2 - k_{nm}^2}\, r\right) J_n^*\left(\sqrt{k_d^2 - k_{nm}^2}\, r\right) r\, dr\right)^{1/2}$ is the normalization factor.

Similar to the discussion in the previous section, if a rigid tube boundary condition is considered, then $g(r,\theta) = J_n(\xi_{nm}r/b)e^{in\theta}$, where, as before, $\xi_{nm}$ represents the $m^{th}$ zero of $J'_n$. The modal wavenumber in this case is $k_{nm} = \sqrt{k_1^2 - (\xi_{nm}/b)^2}$.

Note that if only propagating, non-decaying modes are considered, the summation of Eq. (14) is over a finite number of modes as shown for the axisymmetric case ($n=0$) in Section 2 and in [7,9] for the general case. Each of the traveling waves, $\eta_{nm}^{\pm}$, takes the form of Eq.(10), with the relevant wavenumber, $k$, assigned to the modal wavenumber, $k_{nm}$. As discussed before the relation of each dispersion branch relates the modal wavenumber, $k_{nm}$, to the angular frequency $\omega$.

The set of cross-section modulation functions of Eq. (14), $g_{nm}$, form a finite basis of the pressure thus one may find the projection on all modal waves simultaneously by solving [1]

$$\begin{bmatrix} \langle p, g_1 \rangle \\ \vdots \\ \langle p, g_n \rangle \\ \vdots \\ \langle p, g_N \rangle \end{bmatrix} = \begin{bmatrix} \langle g_1, g_1 \rangle & \cdots & \langle g_1, g_n \rangle & \cdots & \langle g_1, g_N \rangle \\ & \ddots & & & \\ & & \langle g_n, g_n \rangle & & \\ & & & \ddots & \\ & & & & \langle g_1, g_N \rangle \end{bmatrix} \begin{bmatrix} \eta_1 \\ \vdots \\ \eta_n \\ \vdots \\ \eta_N \end{bmatrix}. \tag{17}$$

The inner product is defined over the inner tube cross-section

$$\langle f, g \rangle = \int_{-\pi}^{\pi} \int_0^b f(r,\theta) g^*(r,\theta) r dr d\theta, \tag{18}$$

For brevity, all the admissible mode order pairs are denoted using only subscript $n$. Thus, the pressure field can be decomposed into its modal waves which, are indicators of different energy propagation mechanisms.

Thus, by projecting the pressure field of (14) using Eq. (17) to extract the modal waves, $\eta_n$, one can define the modal traveling wave ratio, TWR$_n$, and modal traveling direction, $\text{dir}_n = \pm 1$, of each modal wave similarly to the single propagating mode case as

$$\text{TWR}_n \equiv 1 - \frac{|P_n^+ - P_n^-|}{P_n^+ + P_n^-}, \quad \text{dir}_n \equiv \text{sign}(P_n^+ - P_n^-). \tag{19}$$

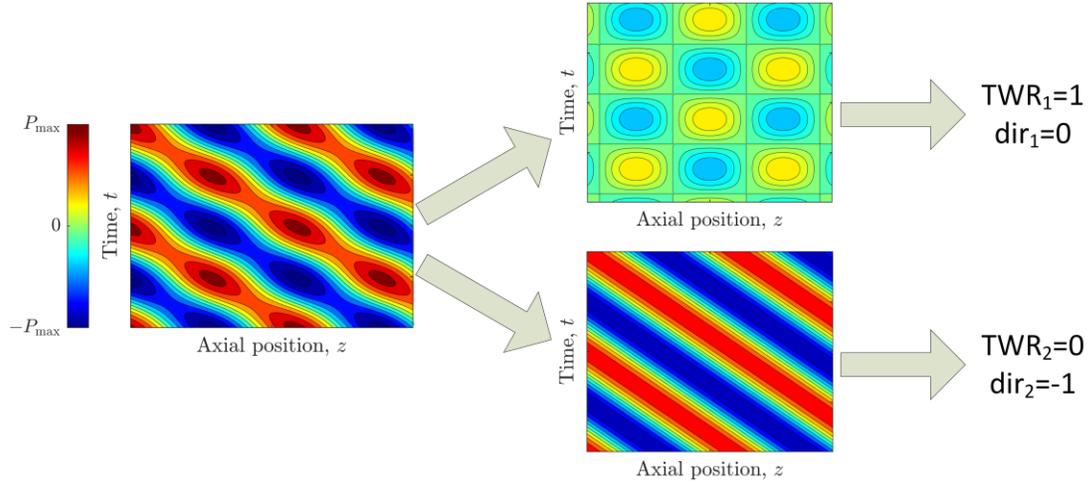

Fig. 3 – (color online) Illustration of the decomposition procedure for the modal traveling wave ratio. The left pressure map shows the measured pressure, which results from the combination of the two propagating modes shown in the right pressure maps; each map is characterized by the modal TWR and dir presented in Eq. (19). The local pressure level is color-coded, neutral (green) stands for zero, warm (red) for positive, cold (blue) for negative pressure fluctuations.

Fig. 3 illustrates the advantage of using wave decomposition and producing the modal TWR. The left pressure map, which is the simulated experiment raw measured one, does not provide a complete picture of the propagation phenomena. Therefore, by decomposing the two propagating modes (pressure maps on the right), it can be seen that the first mode (top) is a pure standing wave ($\text{TWR}_1 = 1$), while the second is a pure backward propagating traveling wave ($\text{TWR}_2 = 0$, $\text{dir}_2 = -1$). Thus, the modal TWR enables one to achieve a better understanding of the propagating phenomena.

## 4. Decomposition and identification of the modal traveling wave ratio

From the illustrated example shown in Fig. 3, it proves advantageous to be able to decompose propagating waves into the physical basis (modes) and quantify each one of them using the modal TWR. Several methods for the experimental modal traveling wave decomposition are outlined in this section. The methods can be applied to the finite-length waveguide dealt with here in the multimode propagation scenario under the following assumptions: (i) $S$ sensors are located along the propagation direction, $z$, their axial locations are denoted as $z_s$, and they share the same position in each axial cross-section ($r, \theta$). (ii) All sensors are sampled simultaneously, where $F_s$ and $\Delta t = F_s^{-1}$ denote the sampling frequency and time accordingly. (iii) The dispersion relation of all modes are known, either from analytical models [1,2] or obtained experimentally [8]. (iv) Propagation attenuation is negligible compared to that at the boundaries [8].

Due to assumption (i), the radial and azimuthal dependency contribution is eliminated, at the cost of losing flexural order's detectability. The pressure wave sampled at location $z_s$ and at time $t_r = r\Delta t$, $r=0,1,\ldots$ is therefore given as:

$$p(b,0,z_s,t_r) = \sum_{n=1}^{N} g_n(b,0)\eta_n(z_s,t_r) = \sum_{n=1}^{N} g_n(b,0)\left(P_n^+ e^{i(\omega t_r - k_n(\omega)z_s)} + P_n^- e^{i(\omega t_r + k_n(\omega)z_s)}\right). \tag{20}$$

The multiplication of each modal wave by a constant does not affect the calculation of the modal TWR or its direction, which is the current effort's purpose. Consequently, these constants can be absorbed into the waves' amplitudes.

By using a Cartesian representation for each wave amplitudes

$$P_n^+ = a_n + ib_n, \quad P_n^- = c_n + id_n, \tag{21}$$

each propagating mode can be written as a vector product

$$p_n(z_s,t_r) = \mathbf{h}_{r,s}^n \mathbf{w}_n. \tag{22}$$

In which $\mathbf{h}$ denotes the wave basis function vector evaluated at the $s$ sensor and $r$ time-step

$$\mathbf{h}_{r,s}^n = \begin{bmatrix} \cos(\omega t_r - k_n z_s) & \sin(\omega t_r - k_n z_s) & \cos(\omega t_r + k_n z_s) & \sin(\omega t_r + k_n z_s) \end{bmatrix}, \tag{23}$$

and $\mathbf{w}_n$ denotes the modal wave constant vector

$$\mathbf{w}_n = \begin{bmatrix} a_n & -b_n & c_n & -d_n \end{bmatrix}^T. \tag{24}$$

Once $\mathbf{w}_n$ is obtained, the corresponding modal TWR can be computed as

$$\text{TWR}_n = 1 - \frac{\left|\sqrt{\mathbf{w}_n^2(1) + \mathbf{w}_n^2(2)} - \sqrt{\mathbf{w}_n^2(3) + \mathbf{w}_n^2(4)}\right|}{\sqrt{\mathbf{w}_n^2(1) + \mathbf{w}_n^2(2)} + \sqrt{\mathbf{w}_n^2(3) + \mathbf{w}_n^2(4)}}, \quad (25)$$

and the modal traveling wave direction can be decided from

$$\text{dir}_n = \text{sign}\left(\sqrt{\mathbf{w}_n^2(1) + \mathbf{w}_n^2(2)} - \sqrt{\mathbf{w}_n^2(3) + \mathbf{w}_n^2(4)}\right), \quad (26)$$

where $\mathbf{w}_n(j)$ denotes the $j^{\text{th}}$ element of the vector $\mathbf{w}_n$ whose coefficients describe the $n^{\text{th}}$ modal propagating wave. Having identified each mode's wave parameter vector is equivalent to the identification of its modal TWR.

Under assumption (iv), the model presented above is accurate when dissipation in the propagation can be negligible, i.e., when the fluid is modeled as a lossless media. This assumption may be used if the liquid viscosity is small compared to the boundary attenuation effects [8]. One may include attenuation effects by the use of a complex wavenumber [9,11] $k_n = \gamma_n - i\alpha_n$, which will alter Eq. (23) to the following form

$$\mathbf{h}_{r,s}^n = \left[\left[\cos(\omega t_r - \gamma_n z_s) \quad \sin(\omega t_r - \gamma_n z_s)\right]e^{-\alpha_n z_s} \quad \left[\cos(\omega t_r + \gamma_n z_s) \quad \sin(\omega t_r + \gamma_n z_s)\right]e^{\alpha_n z_s}\right]. \quad (27)$$

Thus assumption (iv) is not necessary, but from the analysis presented in [8] and as will be shown here, when the fluid under consideration is air, propagation-related attenuation can be neglected without introducing noticeable inaccuracies, and the assumption (iv) holds.

### 4.1 Multichannel least-squares method

The experimental setup requirements in terms of sensor placement and the model for the model-based methods are outlined in this section. Under the assumptions presented at the beginning of section 4, by denoting the pressure measurement of a sensor located at the position $z_s$ at a sampled time $r$ as $p_{r,s}$, given the time history of a batch at $r=1,\ldots, R$ sampled times, and for the $s=1,\ldots, S$ sensors, the modal coefficient vector can be identified by solving the following linear least-squares problem

$$\min_{\mathbf{W}} \|\mathbf{Y} - \mathbf{A}\mathbf{W}\|^2. \quad (28)$$

Here **Y** stands for the measured data vector measured from all sensors. **A** is the model matrix of dimensions $RS \times 4N$, composed of an assembly of $\mathbf{h}_{r,s}^n$ for each sensor and mode, which is built on the basis of the physical modeling, and **W** denotes the linear coefficient vector that defines the various wave amplitudes and modal TWR, their full description can be found in Appendix A.

A solution $\hat{\mathbf{W}}$ to the minimization problem Eq. (28) can be found by ([42])

$$\hat{\mathbf{W}} = \mathbf{A}^\dagger \mathbf{Y}, \tag{29}$$

where $\mathbf{A}^\dagger$ is the left pseudoinverse of **A**, which is computed by known decomposition methods [42]. The solution is unique as long as **A** is uniquely left invertible, i.e., it has full column rank (rank(**A**) = 4$N$) [42].

### 4.1.1 Error analysis of the model inversion via least-squares

When considering the wave decomposition, errors that arise due to noise, uncertainties in geometrical dimensions, sensor spacing, and physical model uncertainties [33] should be accounted for. In this subsection, the solution's existence and accuracy are investigated based on perturbation of the matrix linear system of Eq (29).

Equation (22) can be written as the multiplication of a temporal and spectral matrix as:

$$p_n(z_s, t_r) = [\cos\omega t_r \quad \sin\omega t_r] \mathbf{M}_{ns} \mathbf{\Theta}_n, \tag{30}$$

where $\mathbf{M}_{ns}$ denotes the modal spatial matrix incorporating the sensors' placement,

$$\mathbf{M}_{ns} = \begin{bmatrix} \cos k_n z_s & 0 & 0 & \sin k_n z_s \\ 0 & \cos k_n z_s & \sin k_n z_s & 0 \end{bmatrix}, \tag{31}$$

and the transformed amplitude vector $\mathbf{\Theta}_n$ is defined by the following transformation

$$\mathbf{\Theta}_n = \begin{bmatrix} \begin{bmatrix} 1 & 0 \\ 0 & 1 \end{bmatrix} & \begin{bmatrix} 1 & 0 \\ 0 & 1 \end{bmatrix} \\ \begin{bmatrix} 1 & 0 \\ 0 & -1 \end{bmatrix} & \begin{bmatrix} -1 & 0 \\ 0 & 1 \end{bmatrix} \end{bmatrix} \mathbf{w}_n. \tag{32}$$

Here, the elements of the transformed modal amplitude vector are denoted as

$$\mathbf{\Theta}_n \equiv \begin{bmatrix} A_{cn} & B_{cn} & B_{sn} & A_{sn} \end{bmatrix}^T. \tag{33}$$

The measured pressure at sensor $s$ at all given times can be derived by using the finite modal summation

$$p(z_s, \mathbf{t}) = [\cos \omega \mathbf{t} \quad \sin \omega \mathbf{t}] \mathbf{M}_s \mathbf{\Theta}, \tag{34}$$

where the matrix $\mathbf{M}_s$ is composed of the modal spatial matrices and $\mathbf{\Theta}$ is the $4N \times 1$ transformed modal amplitude vector of all modes. Multiplying Eq. (34) by the generalized left inverse temporal matrix gives

$$\mathbf{M}_s \mathbf{\Theta} = \begin{bmatrix} P_{\cos}(z_s, \omega) \\ P_{\sin}(z_s, \omega) \end{bmatrix} \equiv P_s, \tag{35}$$

where $P_{\cos}$ and $P_{\sin}$ are the real Fourier coefficients of the pressure signal measured at each sensor. Concatenating all sensors' models and the real Fourier coefficients in a similar manner leads to the augmented model

$$[\mathbf{M}]_{2S \times 4N} [\mathbf{\Theta}]_{4N \times 1} = [\mathbf{P}]_{2S \times 1}. \tag{36}$$

Thus, one can identify the Fourier coefficients, and subsequently, the modal wave constants can be found. For M to be uniquely left invertible, it has to have a full column rank, which requires that the row dimension be equal or larger to that of the columns. Therefore, there is a need for at least $4N$ linearly independent rows. This condition is satisfied if there are at least $2N$ sensors and as long as the requirement for sensor placement is fulfilled, i.e., not all sensors are separated by an integer multiplication of half-wavelength. If the latter does not stand, it is guaranteed that at least two columns of $\mathbf{M}$ will be linearly dependent.

The modal amplitudes' error bounds are investigated via a perturbation analysis of the identified modal amplitude vector to uncertainties in the model and measured data. The modal amplitudes perturbation norm is bounded by [43]:

$$\|\delta \mathbf{\Theta}\|_2 \leq \frac{\text{cond}(\mathbf{M})}{1 - \text{cond}(\mathbf{M}) \varepsilon_\mathbf{M}} \left( \|\mathbf{\Theta}\|_2 \varepsilon_\mathbf{M} + \varepsilon_\mathbf{P} \frac{\|\mathbf{P}\|_2}{\|\mathbf{M}\|_2} + \varepsilon_\mathbf{M} \text{cond}(\mathbf{M}) \frac{\|\mathbf{P} - \mathbf{M}\mathbf{\Theta}\|_2}{\|\mathbf{M}\|_2} \right) \tag{37}$$

in which $\delta$ denotes the perturbation, $\varepsilon_\mathbf{M} \geq \|\delta \mathbf{M}\|_2/\|\mathbf{M}\|_2$, and $\varepsilon_\mathbf{P} \geq \|\delta \mathbf{P}\|_2/\|\mathbf{P}\|_2$ are the upper bound on the normalized perturbation of $\mathbf{M}$ and $\mathbf{P}$, and $\text{cond}(\mathbf{M}) = \|\mathbf{M}\|_2 \|\mathbf{M}^\dagger\|_2$ denotes the condition number. To keep the modal amplitude vector perturbation small, the following condition should hold: $\text{cond}(\mathbf{M}) \approx 1$, $\|\mathbf{M}\|_2 \geq 1$, $\varepsilon_\mathbf{p} \ll 1$, and $\varepsilon_\mathbf{M} \ll 1$. If the first condition is satisfied, the second one is automatically satisfied by the definition of the condition number. A comparison between this condition and the classical separation distance of [50–52] is given in appendix B for the case of single-mode and 2 sensors.

In this work, the stepped sine excitation was used, which ensures that the error in the identification of the Fourier's coefficients $\varepsilon_\mathbf{P}$ is kept as small as required [53]. The last condition is satisfied as long as the wavenumbers and sensors' position are known up to a reasonable engineering tolerance, as is shown in the example of section 4.1.2. The sensors' placements dictate the condition number of the matrix, and therefore it is of special importance.

In [35], an optimization procedure was introduced to minimize the model matrix condition number for a wide frequency range. In this research, the sensors were placed such that at 3000 Hz, the condition number is kept small for all models discussed in the following sections. This was achieved by combining separation distances of 0.1 and 0.15 m between the sensors. Moreover, since each sensor may abrupt the waveform and give rise to boundary effects, the sensors are located far away from each other. Therefore, the sensors' placement does not obey the rule that was suggested for the two-microphone impedance tube by [50–52]. Since a full dispersion model is fitted, spatial aliasing does not occur. The advantage of placing each pair of sensors in less than half-wavelength is not clear in the multichannel model inversion and is not adapted in this work. Multichannel allows one to minimize the model matrix condition matrix at a large frequency region, as shown in [35]. Moreover, it allows one to include additional modes to the physical model, as discussed here. Therefore, it is concluded that the multichannel approach is crucial to a successful decomposition of the pressure wave in the case of multimode propagation.

**4.1.2 Error analysis for model uncertainties in the experimental system**

The effect of uncertainty in geometrical dimensions and wavelength on $\varepsilon_\mathbf{M}$ were evaluated numerically using a Monte-Carlo statistical test. The model on which these tests are carried out is shown in the following sections, both in simulations and experiments. Eleven sensors are placed in their nominal locations, as given in Table 2. A model of 3 propagating modes (which will be discussed in section 5) is used. Each nominal sensor position is perturbed using stochastic parameters with a standard deviation $\sigma_p =$ 0.5 mm similar to the assumption used in [51]. Another stochastic parameter is used to determines the uncertainty of each wavenumber. The total standard deviation of the wavelength, $\sigma_k$, was set to 0%, 0.5%, and 1% with respect to the nominal values. The Monte-Carlo test was run 100,000 times, and the histogram for each case of $\varepsilon_\mathbf{M}$ is given in Fig. 4.

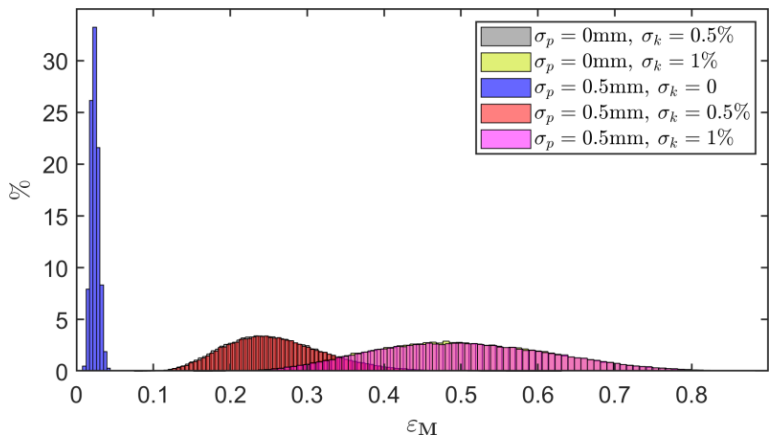

Fig. 4 – (color online) Histogram of the model matrix normalized perturbation results based on a randomized Monte-Carlo test. Sensor positioning standard deviation, $\sigma_p$, and wavenumber uncertainties standard deviations, $\sigma_k$, are given in legend for each histogram.

The histograms emphasize the importance of an accurate dispersion model. The uncertainties in geometrical dimensions do not cause a large error, as long as the wavenumbers are exact. This can be seen from the 2 cases where the microphone positioning is deterministic, and only wavenumbers uncertainties are considered, which are shown in Fig. 4 in gray and yellow (color online). For both of these cases, the histograms are almost covered by these of the same $\sigma_k$ and $\sigma_p = 0.5$ mm (red in front of gray and magenta in front of yellow).

## 4.2 Multichannel recursive-least-squares method

In the cases where either batch measurement is not available or that the inversion of the full model matrix **A** is computationally unfeasible, Eq. (28) minimization problem can be solved by employing the MC-RLS method [37]. The MC-RLS method requires an initialization of the weight vector, $\mathbf{W}_0$, and the information matrix, $\mathbf{P}_0$, which is commonly chosen to be the identity matrix multiplied by a constant scalar, and a prechosen scalar forgetting factor $\lambda$ [37]. It is assumed that the new samples, $p_{r,s}$, are measured simultaneously from all sensors at times $t_r$. The modal wave amplitudes and TWR can be identified by using the traveling wave basis matrix

$$\mathbf{H}_r = \begin{bmatrix} \mathbf{h}_{r,1}^1 & \cdots & \mathbf{h}_{r,1}^N \\ \vdots & \ddots & \vdots \\ \mathbf{h}_{r,S}^1 & \cdots & \mathbf{h}_{r,S}^N \end{bmatrix}_{S \times 4N}, \tag{38}$$

as the multichannel filtered-reference matrix ($\mathbf{V}(n)$ in [37, Eq. (20)]), following the schematic given in [37, Eq. (23)-(25)] for the $r^{\text{th}}$ step, which can be found in Appendix C.

## 4.3 Multichannel least-mean-squares method

Although the RLS method is known to converge faster than the LMS method, in the case of multichannel, the need to invert an $S \times S$ matrix arises at every time step of the RLS algorithm [37]. Thus, it is preferable to use the LMS method, especially in the case of a high sampling rate on fixed-point hardware (FPGA, for example). The MC-LMS method [37] requires initialization of the weight vector, $\mathbf{W}_0$, and a prechosen, small, scalar constant step size $\mu$. The MC-LMS algorithm for the $r^{\text{th}}$ step schematic is given in [37, Eq. (11)–(13)], where again the traveling wave basis matrix $\mathbf{H}_r$ is used instead of the filtered-reference matrix and is also outlined in Appendix D.

The MC-LMS method requires a significantly lower computational effort than the MC-RLS method and can, therefore, be implemented in real-time on a digital processor. The modal traveling wave amplitudes and modal TWR for each of the participating modes can be monitored in real-time.

**4.4 Synchronous demodulation method**

All three estimation methods described above result in an identification error if only a partial propagation model is used, i.e., some of the propagating modes are omitted (as shown in Section 5). To overcome this deficiency, the use of a modified synchronous detection method is proposed. The method uses a nonlinear filter of the following form; consider the signal presented in Eq. (14) under the cross-section sensor location assumption (i), it can be written by using an axial-dependent amplitude of temporal Cosine and Sine functions [17], as shown in Eq. (35)

$$p(z_s, t_r) \equiv p_{s,r} = P_{\cos}(z_s)\cos(\omega t_r) + P_{\sin}(z_s)\sin(\omega t_r), \tag{39}$$

where $P_{\cos}$ and $P_{\sin}$ are in-phase and in-quadrature, respectively. The synchronous demodulation [38] can be used to extract the in-phase and in-quadrature components of a signal with a known angular frequency by multiplying the signal with Cosine and Sine and filtering high harmonics

$$\begin{pmatrix} P_{\cos}(z) \\ P_{\sin}(z) \end{pmatrix} = 2 \cdot \text{LPF}\left\{ \begin{pmatrix} \cos(\omega t) \\ \sin(\omega t) \end{pmatrix} p(z,t) \right\}, \tag{40}$$

in which LPF stands for a real-time low-pass filter with no static gain. Equation. (40) can be implemented directly every time a new measurement is acquired. The amplitudes $P_{\cos}$ and $P_{\sin}$ are both a linear combination of the complex modal amplitude and can be expanded in such a summation, as shown in Eq. (34). To extract a single element from the transformed modal amplitude vector, introduced in Eq. (33), spatial synchronous demodulation can be applied subsequently to the temporal synchronous demodulation

$$\begin{pmatrix} A_{cn} \\ B_{cn} \\ A_{sn} \\ B_{sn} \end{pmatrix} = 2\text{LPF}\left\{ \begin{pmatrix} P_{\cos} & 0 \\ P_{\sin} & 0 \\ 0 & P_{\cos} \\ 0 & P_{\sin} \end{pmatrix} \begin{pmatrix} \cos(k_n z) \\ \sin(k_n z) \end{pmatrix} \right\}. \tag{41}$$

The discrete nature of the LPF is enforced by the use of a finite number of samples and discrete locations of sensors. The modal traveling wave amplitudes and the modal TWR of the participating modes can be computed from $A_{cn}$, $B_{cn}$, $A_{sn}$, and $B_{sn}$ by using the transformation defined in Eq. (32).

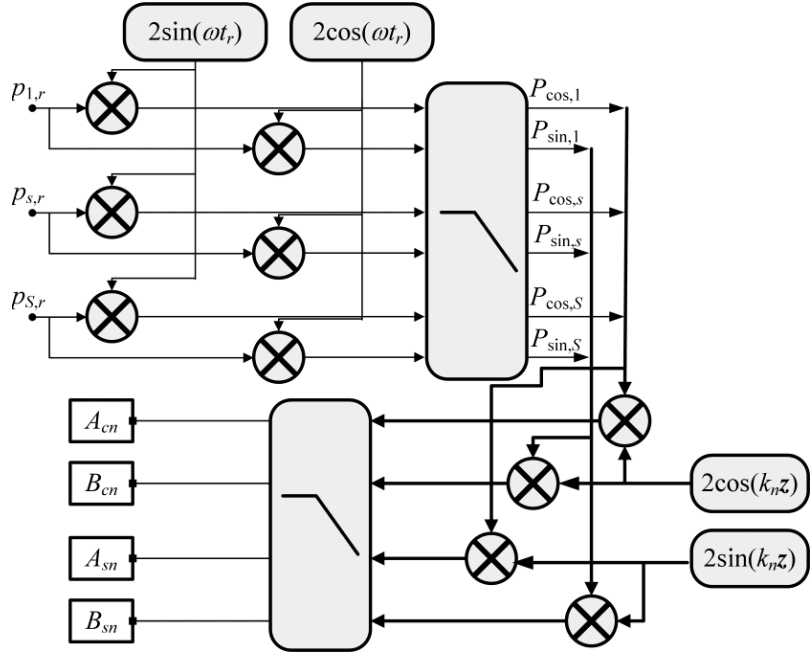

Fig. 5 – Synchronous demodulation method block diagram. Bold lines denote vectors, and the multiplication of vectors is defined element-wise. $z$ denotes the sensor position vector, and the two tall blocks represent the temporal and spatial low-pass filters. The outcome is the modal parameter vectors.

A schematic block diagram of the synchronous demodulation method is presented in Fig. 5. The measured pressure signals are passed through the temporal synchronous demodulation to identify the Fourier's coefficients. Then, the Fourier's coefficient vector is passed through the spatial synchronous demodulation to identify the modal wave amplitudes. Note the difference between the two LPF operations, while the temporal block represents $2S$ channels that are being updated simultaneously at sampling times, the spatial block represents four vectorized channels which are memoryless, meaning that the maximal filter order is equal to the number of sensors. As shown in Fig. 5 and the above discussion, the synchronous demodulation can be used to decompose and identify only the $n^{th}$ mode modal amplitudes and modal TWR without considering additional modes. The method's downfall lies in the need to use a spatial LPF, which is limited by the sensors' number and spatial distribution.

## 5. Sensitivity based analysis to choose the model-order

The number of modes to be included in the wave model of Eq. (14) is not known a priori. Therefore, a sensitivity analysis [45], should be carried out to avoid overfitting [42]. A simulation was created, and on its basis, the proposed sensitivity analysis and its merits are discussed. The simulation was based on a simulated measurement signal (Eq. (14)), which was created using three propagating modes summation (the two asymmetrical modes and a third flexural mode) at an excitation frequency of 3000 Hz. The simulated signal was measured at 11 locations along the $z$-axis at the inner tube radius at the same azimuthal angle. A normally distributed measurement noise model was added to all sensors with a prechosen signal to noise ratio (SNR). The numerical values used in the sensitivity analysis simulation were chosen based on the experimental system properties and measurements and are provided in Table 2, in which the modes are ordered from the most to the least dominant. Note the modal amplitude ratios between the dominant (first) mode and each of the other mode absolute amplitudes, which may intuitively lead to the (wrong) conclusion that the secondary modes may be neglected from the physical model. A model that includes four modes is also considered to evaluate the overfitting effect. Since the simulated signal did not include the fourth mode, its modal amplitudes are considered zeros, as shown in Table 2.

The waves' complex amplitudes are identified using the MC-LS method, where each time, one additional mode was added to the model. The sensitivity of the parameters was analyzed using the condition number of the model matrix **A**, which is an indicator of the sensitivity to small perturbations of the used model [42]; the estimated scaled residual norm

$$\hat{\sigma}^2 = \frac{\hat{\mathbf{r}}^T \hat{\mathbf{r}}}{RS - 4N}, \quad \hat{\mathbf{r}} \equiv \mathbf{Y} - \mathbf{A}\hat{\mathbf{W}}, \tag{42}$$

where $\hat{\mathbf{r}}$ denotes the estimated residual and $\hat{\mathbf{W}}$ is computed from Eq. (29); and the normalized amplitude error

$$e_n = \sqrt{\left(\frac{\hat{P}_n^+ - P_n^+}{P_n^+}\right)^2 + \left(\frac{\hat{P}_n^- - P_n^-}{P_n^-}\right)^2}, \tag{43}$$

in which $\hat{P}_n^\pm$ are the estimated complex amplitudes of the $n^{th}$ mode computed by using $\hat{\mathbf{W}}$, Eq. (24), and Eq. (21).

Table 2 – Numerical values used for the analysis of Sec. 5-7

| Property | Value, units |
| --- | --- |
| Sampling frequency | 250, kHz |
| Excitation frequency | 3000, Hz |
| Number of cycles | 100 |
| Signal to Noise Ratio | 5 (Sec. 5) / ∞ (Sec. 6) |
| Modal wavenumbers | (54.8189,12.1459,21.4010,16.3615), rad/m |
| Forward modal amplitudes | (0.0392-0.0207i,0.0041-0.0038i ,0.0023-0.0021i,0) |
| Backward modal amplitudes | (0.0283+0.0444i,-0.0010+0.0148i,0.0011-0.0023i,0) |
| Sensors z positions | (0.3995,0.5016,0.6019,0.7019,0.8521,1.0031,… 1.1512,1.3003,1.4008,1.5015,1.6020), m |

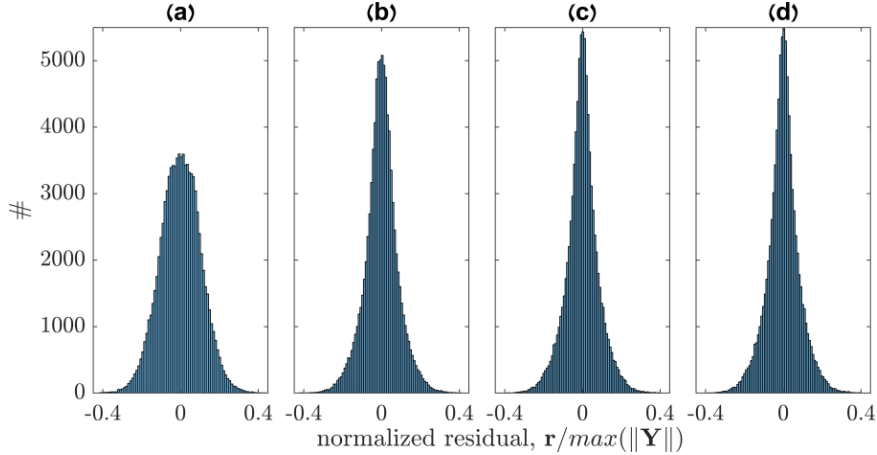

Fig. 6– Simulation normalized residual histogram. Each subplot represents the residual when including the following number of modes in the estimation model: (a) – 1 mode. (b) – 2 modes. (c) – 3 modes. (d) – 4 modes.

Fig. 6 depicts the normalized residual histogram, $\mathbf{r}/max(\|\mathbf{Y}\|)$, when 1 to 4 modes are included in the fitted model. Note that the single-mode model histogram (a) differs from the other three but still seems to have an uncorrelated zero mean nature. The differences between the other three histograms (b, c, and d) are not as notable and do not contribute towards the effort of determining the proper model order. Since the noise level observation is not informative enough to determine the model order, the need to observe the change in another, measurable parameter arises. The estimation's primary goal is to identify the wave complex amplitudes correctly; in other words, one wishes to minimize the normalized amplitude error of a prechosen mode, which is not known a priori. Moreover, when analytical models are used, small perturbations of the models are to be expected. Therefore, if it were possible, one would wish to follow the change in the normalized amplitude error versus the sensitivity of the model to perturbations, which, as discussed in section 4.4.1 (Eq. (37)) is highly dependent on the model matrix condition number, as a function of the model order.

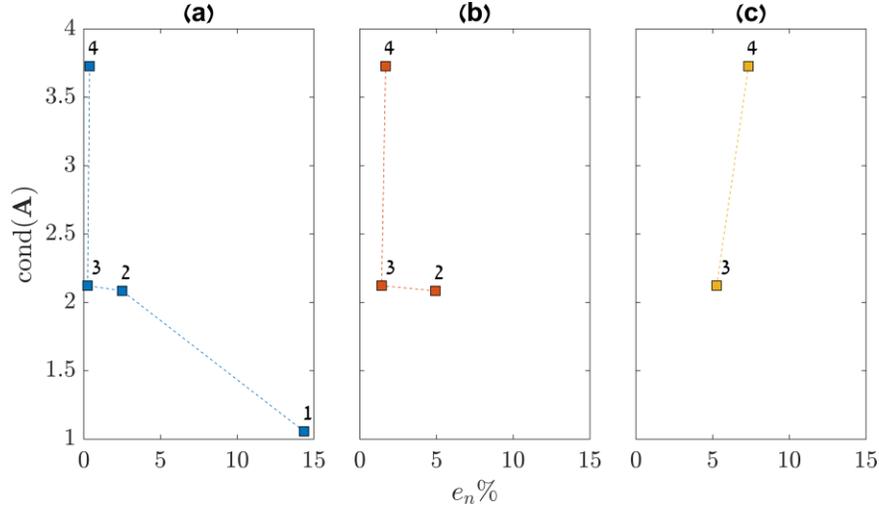

Fig. 7 – (color online) Sensitivity analysis simulation results, showing the model matrix's condition number versus normalized amplitude error (Eq. (43)). Numbers represent the number of modes (Eq. (20)) included in the fitted model (Eq. (28)). Each subplot (a), (b), or (c) is related to the normalized amplitude error of the associated mode: (a) – Dominant mode. (b) – Second dominant mode. (c) – Least dominant mode.

Fig. 7 shows the effect of adding each of the four modes to the model on the condition number and on the normalized estimation amplitude error of the particular mode. Clearly, as can be deduced from Fig.6a, fitting a single-mode model results in a relative error of 15% in the amplitude of this very mode. On the other hand, adding 2,3, and 4 modes to the model creates a clear preferred order when three-modes seem to reduce the fit error without making the model greatly uncertain, as reflected by the sudden increase in condition number upon increasing from 3 to 4 modes in the model. The sudden increase in the condition number is due to the overfitted model; the simulation measured signal was obtained using only the first three modes. Thus, it can be seen that while overfitting will not greatly affect the normalized amplitude error, its effect on the numerical conditioning is nonnegligible and can result in an inaccurate estimation due to large uncertainties of the identified parameters as outlined in section 4.1.1.

The importance of observing both parameters: the model matrix's condition number and the normalized amplitude error parameters together, is clarified. Upon adding a second mode, the dominant mode's fit error is reduced by a factor of 3 to about 5%, while including four modes doubles the model matrix's condition number, thus the sensitivity, without a noticeable effect on the normalized error.

Fig. 6(a) shows the importance of including the correct amount of modes even in the case where only the dominant mode is to be identified; although the dominant mode amplitude is four times larger than that of the second, neglecting additional modes results in a nonnegligible error in the identification.

When considering the more realistic case, in which the exact values are not known and the normalized amplitude error cannot be calculated. A similar approach is proposed by observing the change in the estimated scaled residual norm versus the model matrix's condition number, as a function of the model order, which can be seen in Fig. 8.

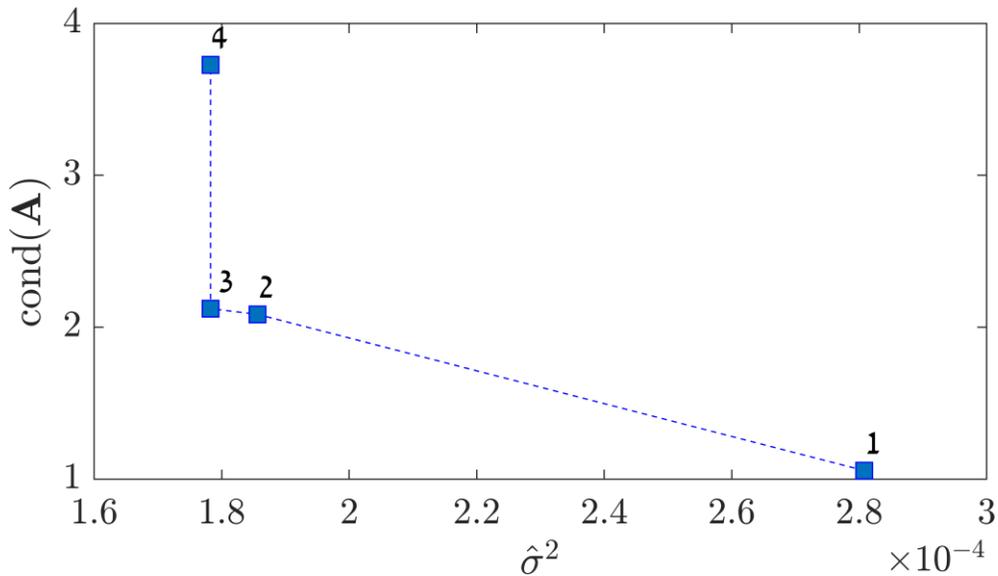

Fig. 8 – Sensitivity analysis simulation result, model matrix's condition number versus estimated scaled residual norm (Eq. (42)). Numbers represent the number of modes (Eq. (20)) that are used in the model (Eq. (28)).

Fig. 8 shows the L-shape-like behavior [45] of adding additional parameters to the LS estimation, from which the optimal number of modes to include can be determined. From this analysis, one can conclude to include three-modes in the model since it resides close to the L-curved "corner" [45,46]. In the current simulation, the two-modes model also resides near the "corner". This is due to the large amplitude ratio between the dominant and the third modes (17/1), which is smaller than the chosen SNR. Still, from the analysis done based on this discussion and Fig. 8 L-curved behavior, the three-modes model is chosen in this case, which corresponds to the correct number of modes used in the simulated signal.

The sensitivity analysis done here applies to the three model-based least-squares methods introduced above; the LS, LMS, and RLS statistical similarities were shown previously in [41].

## 6. Simulated comparison between outlined methods

A numerical comparison between the four outlined methods was performed using the numerical parameters that are provided in Table 2. A model that simulated three modes in the acquired signal, already used in the previous section, is analyzed here, this time without added noise. Fig. 9 shows the results of the simulation-based identified modal TWR for the batch MC-LS method and the three recursive methods. Both model-based recursive methods, MC-RLS and MC-LMS, converged to the batch LS solution for each of the three modes, which is accurate in the absence of noise. The synchronous demodulation was only able to correctly identify the modal TWR associated with the first and third modes. This is due to spatial filter design, which includes a dynamical passband up to the spatial cutoff frequency. For this reason, for the given sensor configuration, the spatial LPF could not reduce the magnitude of the double harmonics [38] of the second mode as required in Eq. (41), resulting in a biased estimation.

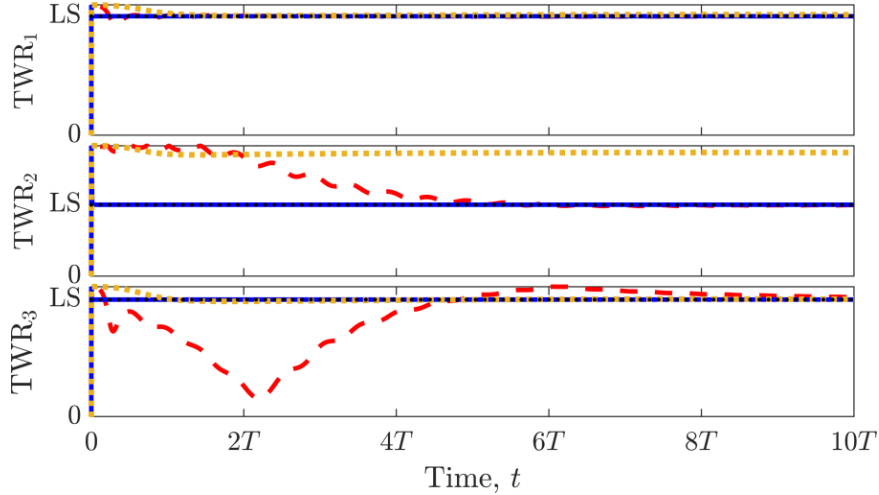

Fig. 9 – (color online) Modal TWR identification simulation results of the outlined methods. LS denotes the MC-LS solution. Each y-axis is bound between 0 to 1. $T$ denotes the simulated signal period. Legend: Dot markers (black) – MC-LS. Solid line (blue) – MC-RLS. Dashed line (red) – MC-LMS. Squares markers (orange) – Synchronous demodulation.

Note that the MC-RLS converges simultaneously to all modal TWR once sufficient data is collected for the noiseless simulation. Based on the discussion presented in Sec. 4.1.1, one can show that the MC-RLS in the absence of noise convergence time is equal to the number of modes used in the model. On the other hand, the MC-LMS convergence rate is slower and differs between each of the modes. The MC-LMS convergence rate speed can be associated with the mode's dominancy; the convergence rate is faster for the more dominant mode.

The difference in convergence time between the model-based methods arises from these methods' nature, especially when noise is negligible. This is since the MC-LMS step-size is tuned at the initialization step for the worst-case scenario, while the MC-RLS step-size is being updated at each step. The MC-LMS convergence time is faster for more dominant modes, from the results shown in Fig. 9, the MC-LMS converges to the most dominant mode in under a cycle. Thus, if the goal is to identify the dominant (first) mode TWR accurately, the MC-LMS is preferred over the MC-RLS due to its ease of implementation and low computational cost.

# 7. Experimental study

The four methods outlined in Sec. 4 were analyzed and compared on an experimental PMMA air-filled wave tube. The comparison was conducted at an excitation frequency of 3000 Hz. In [8], five different dispersion branches were identified at the surrounding frequency region; two of them were identified as the axisymmetric modes, the third was identified as the Timoshenko's flexural mode the remaining two were not analytically identified. Not all five modes were identified at all frequencies considered in [8], emphasizing the need for a model order decision test as suggested above. The section opens with a description of the experimental system followed by a sensitivity analysis of the number of modes on the MC-LS method (similar to that of Sec. 5). Then the simulation discussed in Sec. 6 is done on the experimentally acquired signals. Finally, the experimental identified modal TWR results of the recursive MC-LMS method implemented on a field-programmable gate array (FPGA) in real-time are compared both qualitatively and quantitatively with these of the batch MC-LS method applied to signals measured using a NI[©] data-acquisition device.

## 7.1 Experimental system layout

Fig. 10 presents a photo of the experimental air-filled wave tube. The tube length is 2 m, and the outer diameter is 0.09 m. It is made from PMMA. A 3", 4 Ω, and two 60 W Dayton Audio[©] PC83-4 loudspeaker is connected to both ends of the wave tube. Eleven 9.7 mm diameter omnidirectional Adafruit[©] Max9814 microphones are placed along the tube's inner surface at $\theta = 0$. All microphones were calibrated to reduce amplitude and phase mismatch between them. A current amplifier drives the loudspeakers to eliminate the effect of electrical impedance coupling with the acoustic field.

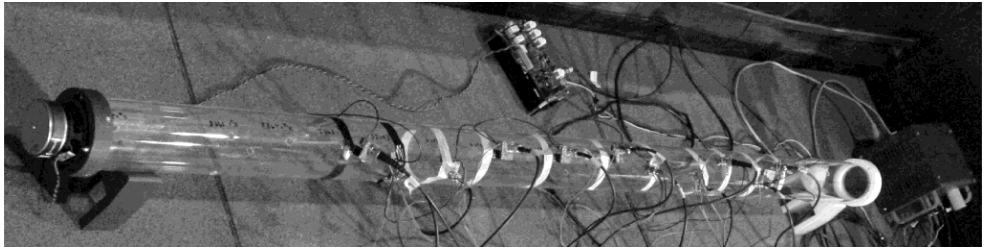

Fig. 10 – Experimental system: an air-filled PMMA cylindrical tube with two Dayton Audio© PC83-4 loudspeakers at each end and eleven Adafruit© Max9814 microphones along the tube axis.

In Table 1, the experimental setup properties are presented. The air density, $\rho_1$, and intrinsic speed of sound, $c_1$, were calculated for a temperature of 25 °C and pressure of 101325 Pa. The PMMA density, $\rho_s$, and Lamé parameters $\lambda$ and $\mu$, were estimated from a set of preliminary experiments.

The batch data used in Sec. 7.2-4 were sampled using a NI© PXIe-6358 data acquisition device and two BNC-2110 connector blocks with 16 analog input channels and a 250 kHz sampling frequency. The recursive MC-LMS was implemented on a dSpace© DS 1005 computation node connected to two DS5203 FPGA boards with 16 analog input channels and an internal clock of 100 MHz that is also used to sample the identified modal TWR at a 10 kHz sampling frequency. The acoustic tube's dispersion curves were identified using the two-actuator phase-perturbations (TAPP) method [8] using a similar setup to the one described here and the NI© data acquisition module.

## 7.2 Multichannel least-squares sensitivity analysis

In the dispersion curves identification phase [8], a total of five propagating modes were identified in the neighboring frequency range. To avoid overfitting and choose the number of modes to include in the TWR identification phase, a sensitivity-based analysis procedure, similar to the one presented in Sec. 5, was carried out.

The last 100 cycles of each microphone's measured signal were used in the sensitivity analysis; the analysis was done using the raw signals where the only filter used is a DC removal high-pass filter to account for the DC-biased voltage outputs of the microphones.

Fig. 11 depicts the experiment normalized residual histogram, $\mathbf{r}/max(\|\mathbf{Y}\|)$, when one to four modes are used in the fitted model. The fifth mode was not considered since its wavenumber is almost identical to the third mode one. This indicates that the third and fifth wavenumbers differ due to split caused by an imperfect circular cross-section; in the discussed sensor placement, the model cannot separate them. Future work should consider adding azimuthal variance between the sensors similar to the setup used in [35] if the modes are to be separable. All four histograms of Fig. 11 have a similar nature. The single-mode model histogram (a) has a higher peak and a wider distribution. As discussed in Sec. 5, it is clear from Fig. 11, the normalized residual by itself cannot provide a clear indication for the number of modes to include in the model.

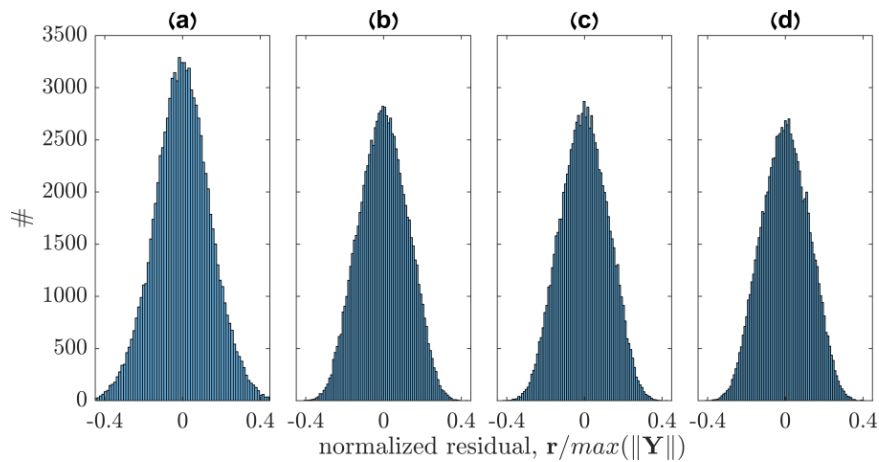

Fig. 11 – Experiment normalized residual histogram. Each subplot represents the residual when including the following number of modes in the estimation model: (a) – 1 mode. (b) – 2 modes. (c) – 3 modes. (d) – 4 modes.

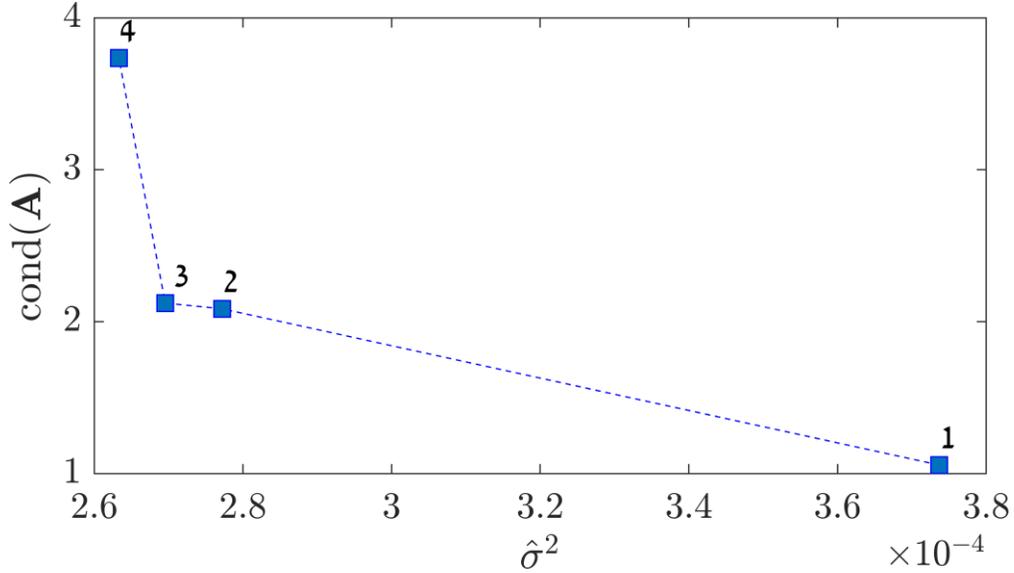

Fig. 12 – Sensitivity analysis experimental result, model matrix's condition number versus estimated scaled residual norm (Eq. (42)). Numbers represent the number of modes (Eq. (20)) that are used in the model (Eq. (28)).

Fig. 12 shows the sensitivity analysis introduced in Sec. 5 performed on the experimentally measured signals from the 11 microphones. The L-shape-like behavior [45,46] is noticeable, and the behavior seems to agree strongly with that of the simulation sensitivity analysis presented in Fig. 8. The resemblance between the simulated and experimental results, Fig. 8 and Fig. 12, can be accounted for by the physical parameters' accuracy; these can be found in Table 2. The main difference between the simulated and experimentally measured signals is that in the simulated signal, the "exact" parameter values and the noise model are known a priori.

It should be noted that in both the experimental result (Fig. 12) and the simulated one (Fig. 8), the addition of a fourth mode to the model caused a reduction in the estimated scaled residual norm. Still, the fourth mode's addition has a significant effect on the model matrix's condition number; it affects the sensitivity to perturbations greatly [42]. Using the L-shape "corner" location, it is concluded that only the first three modes contribute to the acoustic wave-phenomena at this excitation frequency [45].

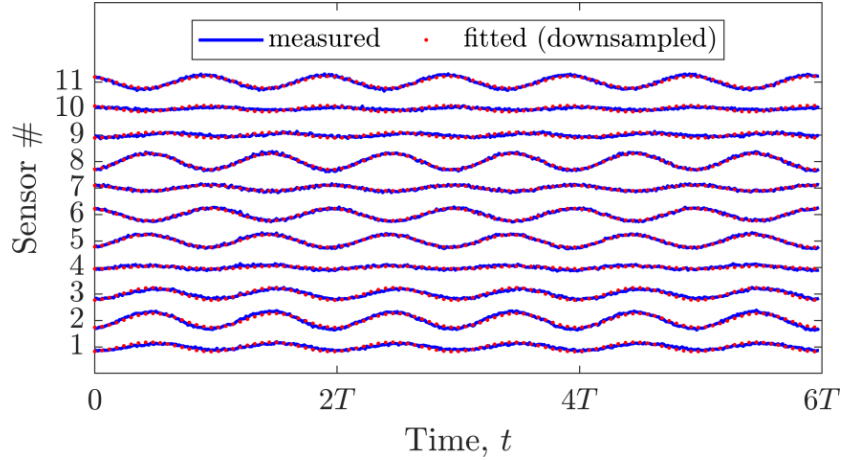

Fig. 13 – (color online) Experimentally measured signal (solid lines) at each sensor and fitted signal (dot markers) using a three-mode model as an input of the MC-LS method. $T$ denotes the excitation signal period time.

Fig. 13 shows the experimentally measured signals of each microphone and the fitted signal based on the modal model of Eq. (22). The decomposition of the modes was done using a 3-mode model and the batch MC-LS method. Clearly, the curve-fitting seems adequate.

From this sensitivity analysis done on the experimentally measured signals, it was concluded that only three modes would be included in the model-based method when decomposing to propagating modes and identifying each mode modal TWR as presented in the following subsections.

### 7.3 Offline numerical comparison of the four methods based on experimentally measured data

To compare the outlined methods (Sec. 4), an offline recursive procedure was realized similarly to that presented in Sec. 6. The input data to the recursive procedure was obtained experimentally from the air-filled acoustic wave tube (Fig. 10), using the NI© board with a similar configuration to the one detailed at the beginning of the previous subsection (Sec. 7.2). The recursive numerical procedure was run offline using Matlab©, where all model-based methods used the three-modes model as was concluded from the experimentally obtained data sensitivity analysis.

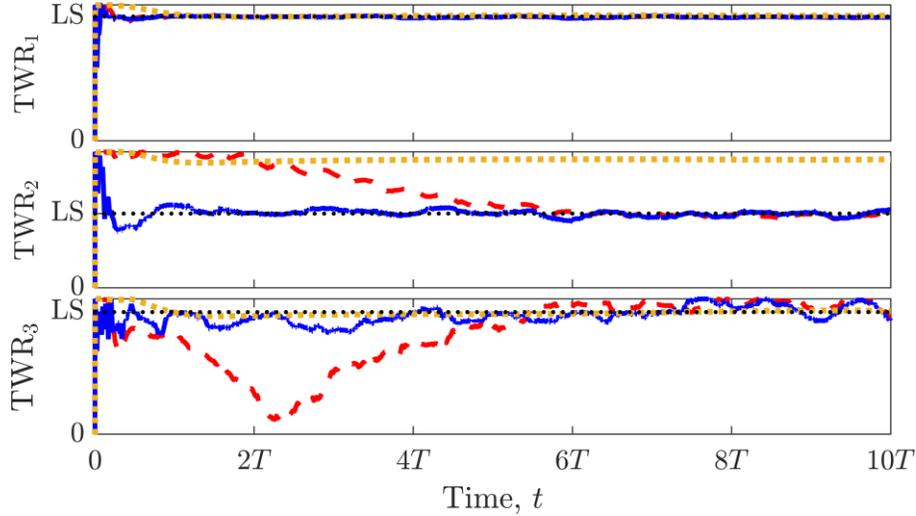

Fig. 14 – (color online) Modal TWR identification experimental results of the four outlined methods. LS denotes the MC-LS solution. Each y-axis is bound between 0 to 1. $T$ denotes the excitation signal period. Legend: Dot markers (black) – MC-LS. Solid lines (blue) – MC-RLS. Dashed lines (red) – MC-LMS. Squares marker (orange) – Synchronous demodulation.

Fig. 14 shows the three identified modal TWR; similarly to the simulation base analysis (Sec. 6), both the MC-RLS and the MC-LMS converged to the least-squares solution for all three modes, and the synchronous demodulation did not converge for the second mode modal TWR. The main difference between the experimentally based analysis (Fig. 14) and the simulated one (Fig. 9) is the presence of noise, which has a significant effect on the MC-RLS convergence time, while its effect on the MC-LMS is less obvious. Thus, when noise is nonnegligible, its magnitude affects the MC-RLS convergence rate advantage, diminishing its advantage, particularly when the computational effort is considered. The latter becomes especially important when implemented on an FPGA. For the MC-LMS, the noise affects the convergence of the weaker modes, but as shown in Fig. 14 and Fig. 9, the MC-LMS converges to the dominant mode modal TWR in under a cycle. The synchronous demodulation's behavior presented, as shown in Fig. 14, is similar to the one of the simulation-based analysis presented in Fig. 9. The presence of noise and its effect on the LPF output is clearly visible, causing weak oscillations around the mean steady-state identified modal TWR.

## 7.4 Comparison between recursive multichannel least-mean-squares and batch multichannel least-squares methods

The recursive MC-LMS was implemented using the two dSpace© DS5203 FPGA boards. A comparison between the batch MC-LS and the recursive MC-LMS experimentally produced identified modal TWR contour maps [18,24], scanned for excitation frequency of 3000 Hz, was conducted. These maps show the effect of amplitude ratio and phase difference of the two loudspeakers (see Fig.10) on the TWR. Clearly, the ratio of traveling and standing waves varies for specific combinations, and the TWR produces this information [18].

Fig. 15 depicts the modal TWR maps results for both methods, left column for the batch MC-LS and right column for the recursive MC-LMS, and for each of the three modes used in the model, top and bottom rows present the results for the most and least dominant modes accordingly. The blue (online) lines of each contour represent the traveling wave regions (TWR=0), while the red lines represent the standing wave region (TWR=1). Note that at the phase shift region of interest, there are two traveling wave regions for each mode. One of them corresponds to the forward and the other to the backward traveling waves.

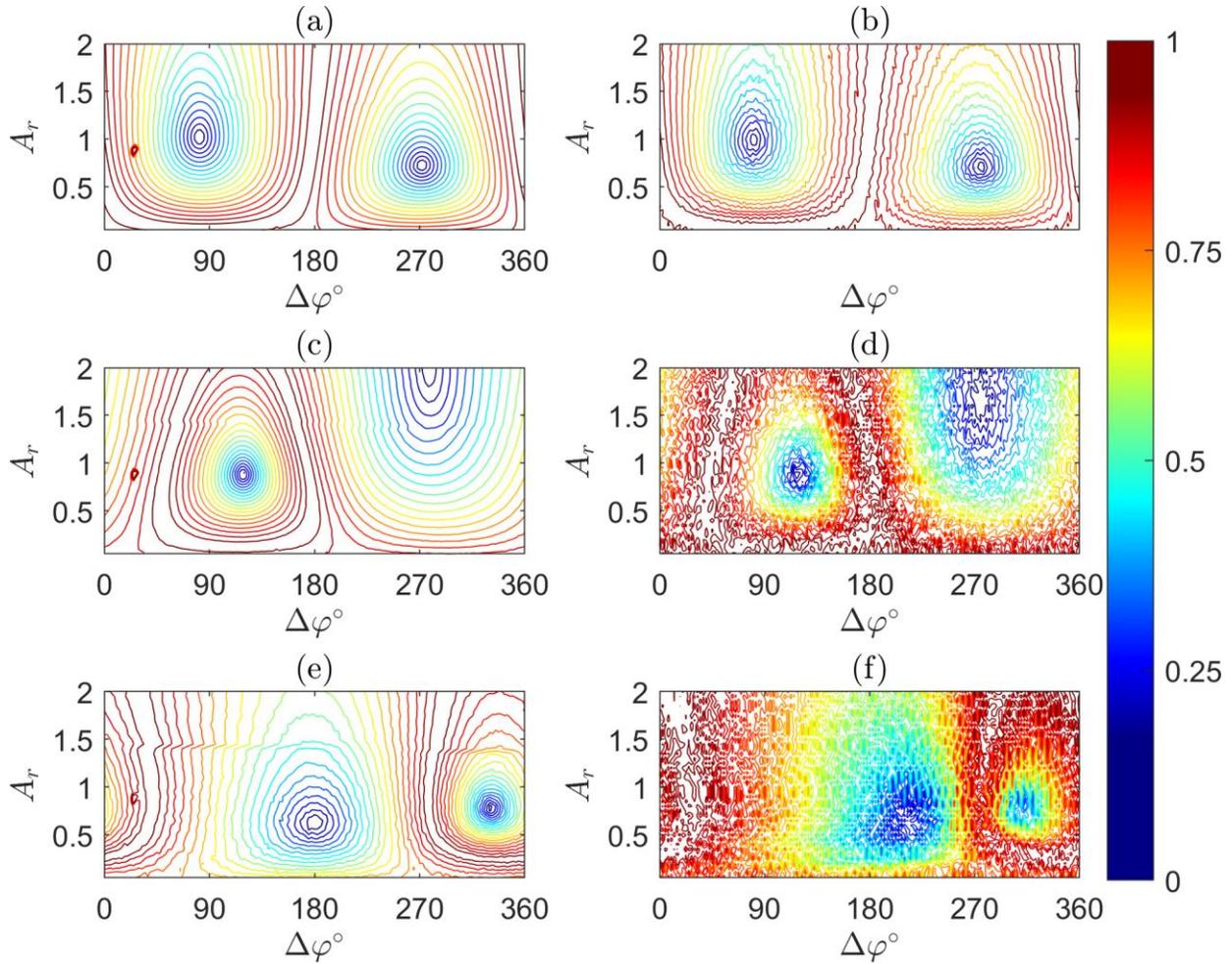

Fig. 15 – (color online) MC-LS and MC-LMS TWR$_n$ contours comparison at 3000 Hz versus loudspeakers amplitude ratio, $A_r$, and phase shift, $\Delta\varphi$, color represent the modal TWR$_n$ values: Blue – 0. Red – 1. (a), (c), and (e) – MC-LS results for the first, second, and third dominant modes. (b), (d), and (f) – MC-LMS results for the first, second, and third dominant modes.

From the modal TWR, the propagation pattern is clear, and if it is desired to convey energy in a chosen direction (TWR=0) or suppress it (TWR=1), it can be done by tuning one of the loudspeakers amplitude and phase relative to the other to achieve the desired modal TWR. The MC-LS obtained contour maps show two anomalies, the first near $A_r = 0.9$ and $\Delta\varphi = 30°$, the second at $A_r = 1.5$ and all phase shifts. These are due to the experimental nature of the identification process. The first may be accounted for by some high-level correlated noise at the frequency of the experiments (external noise); the second anomaly is more subtle, and it may be associated with a small variation in the resistivity of the sensors as a function of time.

When comparing the two methods, one can note that the dominant (first) mode two maps (Fig. 15a and Fig. 15b) are qualitatively almost identical. When comparing the less dominant modes, the MC-LMS modal TWR contours exhibit some distortions, but in principle, one identifies the different regions of the modal TWR (standing, traveling). This may be due to the numerical truncation caused by the finite number of bits in the FPGA realization and due to the finite settling time given for the method at each amplitude ratio and phase shift. The modal TWR contours were also quantitively compared using the relative error norm

$$\varepsilon_n \equiv \frac{1}{\#(A_r)\#(\Delta\varphi)} \left\| \frac{TWR_{n,\text{MC-LS}}(A_r,\Delta\varphi) - TWR_{n,\text{MC-LMS}}(A_r,\Delta\varphi)}{TWR_{n,\text{MC-LS}}(A_r,\Delta\varphi)} \right\|_f, \tag{44}$$

where $A_r$ and $\Delta\varphi$ denote the amplitude ratio and phase shift of the loudspeakers, # denotes the number of elements scanned for each of them, and $\| \ \|_f$ denotes the matrix Frobenius norm. The following results are obtained when applying Eq. (44) to the identified contours of Fig. 15:

$$\varepsilon_1 = 0.08\%, \quad \varepsilon_2 = 0.64\%, \quad \varepsilon_3 = 0.83\%. \tag{45}$$

The relative error norms of each mode show quantitively the same behavior of its qualitative counterpart discussed above. Moreover, the ratios between the dominant mode's relative error norm and each of the less dominant modes are smaller than their associated amplitude ratios. Although $\varepsilon_2$ and $\varepsilon_3$ are only one order of magnitude larger than $\varepsilon_1$ their contour maps seem to have large distortion, but by using a smoothening filter (interpolation), the nature of the modal TWR of these maps can easily be retrieved, as shown in Fig. 15. Thus, the MC-LMS may be used for the real-time identification of the modal TWR and modal amplitudes of all modes.

A more sophisticated LMS algorithm can be used. The interested reader is referred to [44] chapter 3.4 for a comprehensive overview of the many variations. However, their implementation is beyond the scope of this current work, which has shown the feasibility of using the MC-LMS for the real-time decomposition of the pressure fields into its backward and forward traveling wave basis and discussed its advantages in the suggested form.

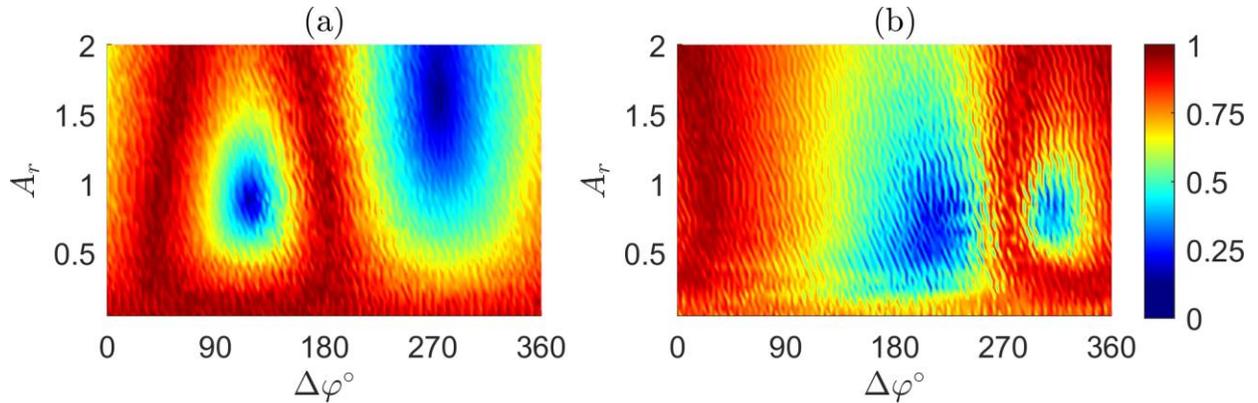

Fig. 16 – (color online) MC-LMS interpolated modal TWR maps, based on the modal TWR contour of Fig. 15, color represents the modal $TWR_n$ values: Blue – 0. Red – 1. (a) – The 2$^{nd}$ mode, (b) The 3$^{rd}$ mode.

## 8. Conclusions

It has been demonstrated that the modal TWR is an effective measure to assess the quality of modal decomposition in the case of an acoustical wave tube. It allows one to understand the wave dynamics better, and it enables one to precisely control the desired propagation pattern by tuning the modal TWR using a secondary transducer. The prospect of using the MC-LMS for a real-time decomposition and identification of the modal amplitudes and the modal TWR was demonstrated experimentally with good agreement to simulation and models.

The importance of including all dominant modes and neglecting others was shown to improve the accuracy and reduce the sensitivity to model uncertainties of the identified modes. The number of modes one should include in the propagation model can be determined using the proposed sensitivity analysis resembling the L-curve method. Methods that do not depend on the full dispersion model, i.e., the number of propagating modes and their wavenumbers, as the synchronous demodulation, require convoluted signal processing analysis and subject to sampling criteria. Thus, model-based methods are preferable in many cases. This is true, especially if one can rely on analytical models or combined analytical and experimental ones for the derivation of the dispersion model.

The MC-LMS has a slower convergence rate compared to the MC-RLS. Still, it is favorable in most cases since it is easy to implement and required a low computational effort. The MC-LMS's merit becomes more notable when noise is introduced to the system.

Hybrid methods like the synchronous demodulation – MC-LMS are a viable option; the first is effective to decompose the temporal harmonics by using time Fourier analysis, the second to decompose the spatial harmonics by using a model-based approach.

Standardized methods to conduct NDT of acoustical properties that use a two-microphone impedance tube and considers a single mode in their physical models are bound to have a biased estimation. As shown in the error analysis (section 4.1.1), these method's accuracy can be enhanced by accounting for the additional modes and using multichannel wave decomposition.

## Appendix A – Equation (28) nomenclature full form

$$\mathbf{A} = \begin{bmatrix} \mathbf{h}_{1,1}^1 & \cdots & \cdots & \cdots & \mathbf{h}_{1,1}^N \\ \vdots & & & & \vdots \\ \mathbf{h}_{R,1}^1 & \cdots & \cdots & \cdots & \mathbf{h}_{R,1}^N \\ \vdots & & & & \vdots \\ \mathbf{h}_{r,s}^1 & \cdots & \mathbf{h}_{r,s}^n & \cdots & \mathbf{h}_{r,s}^N \\ \vdots & & & & \vdots \\ \mathbf{h}_{1,S}^1 & \cdots & \cdots & \cdots & \mathbf{h}_{1,S}^N \\ \vdots & & & & \\ \mathbf{h}_{R,S}^1 & & & & \mathbf{h}_{R,S}^N \end{bmatrix}_{RS \times 4N}, \quad \mathbf{Y} = \begin{bmatrix} p_{1,1} \\ \vdots \\ p_{R,1} \\ \vdots \\ p_{r,s} \\ \vdots \\ p_{1,S} \\ \vdots \\ p_{R,S} \end{bmatrix}_{RS \times 1}, \quad \mathbf{W} = \begin{bmatrix} \mathbf{w}_1 \\ \vdots \\ \mathbf{w}_n \\ \vdots \\ \mathbf{w}_N \end{bmatrix}_{4N \times 1}. \quad (A.1)$$

# Appendix B– Error analysis for a single-mode and two sensors

Following the discussion of section 4.1.1, the case of a two-microphone (rigid) impedance tube model is shown here, and similarities to classical works of [50–52] are drawn.

The model matrix, **M,** in this case, is given by:

$$\mathbf{M} = \begin{bmatrix} \cos k_1 z_1 & 0 & 0 & \sin k_1 z_1 \\ 0 & \cos k_1 z_1 & \sin k_1 z_1 & 0 \\ \cos k_1 z_2 & 0 & 0 & \sin k_1 z_2 \\ 0 & \cos k_1 z_2 & \sin k_1 z_2 & 0 \end{bmatrix}. \tag{B.1}$$

Its singular values are:

$$\sigma_{1,2} = 1 - |\cos k_1 \Delta z|, \quad \sigma_{3,4} = 1 + |\cos k_1 \Delta z|, \tag{B.2}$$

and its condition number is given by

$$\mathrm{cond}(\mathbf{M}) = \frac{\max \sigma}{\min \sigma} = \frac{1 + |\cos k_1 \Delta z|}{1 - |\cos k_1 \Delta z|}. \tag{B.3}$$

Therefore, the condition number tends to infinity whenever

$$\Delta z = \frac{\pi}{k_1} n = \frac{\lambda}{2} n, \quad n \in \mathbb{Z}. \tag{B.4}$$

where $\lambda = 2\pi/k$ is the wavelength.

It should be mentioned that sensor spacing $\Delta z > \lambda/2$ can still render (B.3) unity and is useful when $k_1$ is known in advance.

To conclude, the suggested formulation has recovered the classical results stated in [50–52] for the two-microphone impedance tube while allowing one to check the sensitivity of sensor positioning in the case of multichannel sampling.

# Appendix C – The multichannel recursive-least-squares algorithm [37 Eq. (23)–(25)]

1. Estimate error:

$$\boldsymbol{\alpha}_r = \mathbf{p}_r - \mathbf{H}_r \mathbf{W}_{r-1}, \tag{C.1}$$

2. compute the RLS gain vector:

$$\mathbf{K}_r = \mathbf{P}_{r-1} \mathbf{H}_r^T \left( \lambda \mathbf{I}_S + \mathbf{H}_r \mathbf{P}_{r-1} \mathbf{H}_r^T \right)^{-1}, \tag{C.2}$$

3. update weight vector and information matrix:

$$\mathbf{W}_r = \mathbf{W}_{r-1} + \mathbf{K}_r \boldsymbol{\alpha}_r, \tag{C.3}$$

$$\mathbf{P}_r = \frac{1}{\lambda} \left( \mathbf{I}_{4N} - \lambda^{-1} \mathbf{K}_r \mathbf{H}_r \right) \mathbf{P}_{r-1}, \tag{C.4}$$

where $\mathbf{I}_S$ and $\mathbf{I}_{4N}$ denote the $S \times S$ and $4N \times 4N$ identity matrix.

# Appendix D – The multichannel least-mean-squares algorithm [37 Eq. (11)–(13)]

1. Estimate the equation error:

$$\boldsymbol{\alpha}_r = \mathbf{p}_r - \mathbf{H}_r \mathbf{W}_{r-1}, \tag{D.1}$$

2. update weight vector:

$$\mathbf{W}_r = \mathbf{W}_{r-1} + \frac{\beta}{\|\mathbf{H}_0\|} \mathbf{H}_r^T \boldsymbol{\alpha}_r. \tag{D.2}$$

# Funding

This research did not receive any specific grant from funding agencies in the public, commercial, or not-for-profit sectors.

# References

[1] DEL GROSSO VA, Analysis of multimode acoustic propagation in liquid cylinders with realistic boundary conditions. Application to sound speed and absorption measurements, Acustica. 24 (1971) 299–311.

[2] L.D. Lafleur, F.D. Shields, Low-frequency propagation modes in a liquid-filled elastic tube waveguide, Journal of the Acoustical Society of America. 97 (1995) 1435–1445. https://doi.org/10.1121/1.412981.

[3] R.W. Morse, The velocity of compressional waves in rods of rectangular cross section, Journal of the Acoustical Society of America. (1950). https://doi.org/10.1121/1.1906592.

[4] ASTM International. E1050-12 Standard test method for impedance and absorption of acoustical materials using a tube, two microphones and a digital frequency analysis system. West Conshohocken, PA; ASTM International, 2012. https://doi.org/10.1520/E1050-12

[5] ASTM International. E2611-19 Standard test method for normal incidence determination of porous material acoustical properties based on the transfer matrix method. West Conshohocken, PA; ASTM International, 2019. https://doi.org/10.1520/E2611-19

[6] J. P. Dalmont, Acoustic impedance measurement, part I: A review, Journal of Sound and Vibration, 243.3 (2001) 427-439. https://doi.org/10.1006/jsvi.2000.3428

[7] H. Sato, M. Lebedev, J. Akedo, Theoretical and experimental investigation of propagation of guide waves in cylindrical pipe filled with fluid, Japanese Journal of Applied Physics, Part 1: Regular Papers and Short Notes and Review Papers. 45 (2006) 4573–4576. https://doi.org/10.1143/JJAP.45.4573.

[8] Y. Vered, R. Gabai, I. Bucher, Waveguide dispersion curves identification at low-frequency using two actuators and phase perturbations, The Journal of the Acoustical Society of America. 146 (2019) 2443–2451. https://doi.org/10.1121/1.5128482.

[9] K. Baik, J. Jiang, T.G. Leighton, Acoustic attenuation, phase and group velocities in liquid-filled pipes: Theory, experiment, and examples of water and mercury, The Journal of the Acoustical Society of America. 128 (2010) 2610–2624. https://doi.org/10.1121/1.3495943.

[10] W. Sachse, Y.H. Pao, On the determination of phase and group velocities of dispersive waves in solids, Journal of Applied Physics. 49 (1978) 4320–4327. https://doi.org/10.1063/1.325484.

[11] M.G. Jones, T.L. Parrott, Evaluation of a multi-point method for determining acoustic impedance, Mechanical Systems and Signal Processing. 3 (1989) 15–35. https://doi.org/10.1016/0888-3270(89)90020-4.

[12] M. Oblak, M. Pirnat, M. Boltežar, An impedance tube submerged in a liquid for the low-frequency transmission-loss measurement of a porous material, Applied Acoustics. 139 (2018) 203–212. https://doi.org/10.1016/j.apacoust.2018.04.014.

[13] R.S. Langley, On the modal density and energy flow characteristics of periodic structures, Journal of Sound and Vibration. 172 (1994) 491–511. https://doi.org/10.1006/jsvi.1994.1191.

[14] L. Feng, Acoustic properties of fluid-filled elastic pipes, Journal of Sound and Vibration. 176 (1994) 399–413. https://doi.org/10.1006/jsvi.1994.1384.

[15] A. Søe-Knudsen, S.V. Sorokin, Modelling of linear wave propagation in spatial fluid filled pipe systems consisting of elastic curved and straight elements, Journal of Sound and Vibration. 329 (2010) 5116–5146. https://doi.org/10.1016/j.jsv.2010.06.015.

[16] Y. Vered, I. Bucher, Tailoring phononic-like topologies for controlling the structural-acoustic coupling in fluid-filled cylinders, in: Proceedings of ISMA 2018 - International Conference on Noise and Vibration Engineering and USD 2018 - International Conference on Uncertainty in Structural Dynamics, 2018: pp. 3033–3045.

[17] I. Bucher, Estimating the ratio between travelling and standing vibration waves under non-stationary conditions, Journal of Sound and Vibration. 270 (2004) 341–359. https://doi.org/10.1016/S0022-460X(03)00539-X.

[18] R. Gabai, I. Bucher, Excitation and sensing of multiple vibrating traveling waves in one-dimensional structures, Journal of Sound and Vibration. 319 (2009) 406–425. https://doi.org/10.1016/j.jsv.2008.06.013.

[19] R. Gabai, D. Ilssar, R. Shaham, N. Cohen, I. Bucher, A rotational traveling wave based levitation device – Modelling, design, and control, Sensors and Actuators, A: Physical. 255 (2017) 34–45. https://doi.org/10.1016/j.sna.2016.12.016.

[20] R.A. Scott, An apparatus for accurate measurement of the acoustic impedance of sound-absorbing materials, Proceedings of the Physical Society. 58 (1946). https://doi.org/10.1088/0959-5309/58/3/303.

[21] I. Bucher, R. Gabai, H. Plat, A. Dolev, E. Setter, Experimental travelling waves identification in mechanical structures, Mathematics and Mechanics of Solids. 24 (2019) 152–167. https://doi.org/10.1177/1081286517732825.

[22] AH Von Flotow, Traveling wave control for large spacecraft structures, Journal of Guidance, Control, and Dynamics. 9 (1986) 462-468. https://doi.org/10.2514/3.20133

[23] B.R. Mace, Active control of flexural vibrations, Journal of Sound and Vibration. 114 (1987) 253–270. https://doi.org/10.1016/S0022-460X(87)80152-9.

[24] A. Minikes, R. Gabay, I. Bucher, M. Feldman, On the sensing and tuning of progressive structural vibration waves, IEEE Transactions on Ultrasonics, Ferroelectrics, and Frequency Control. 52 (2005) 1565–1575. https://doi.org/10.1109/TUFFC.2005.1516029.

[25] F. Giraud, C. Giraud-Audine, M. Amberg, B. Lemaire-Semail, Vector control method applied to a traveling wave in a finite beam, IEEE Transactions on Ultrasonics, Ferroelectrics, and Frequency Control. 61 (2014) 147–158. https://doi.org/10.1109/TUFFC.2014.6689782.

[26] W.J. O'Connor, M. Zhu, Boundary-controlled travelling and standing waves in cascaded lumped systems, Mechanical Systems and Signal Processing. 39 (2013) 119–128. https://doi.org/10.1016/j.ymssp.2012.02.005.

[27] H. Habibi, W. O'Connor, Wave-based control of planar motion of beam-like mass–spring arrays, Wave Motion. 72 (2017) 317–330. https://doi.org/10.1016/j.wavemoti.2017.04.002.

[28] I. Peled, W.J. O'Connor, Y. Halevi, On the relationship between wave based control, absolute vibration suppression and input shaping, Mechanical Systems and Signal Processing. 39 (2013) 80–90. https://doi.org/10.1016/j.ymssp.2012.06.006.

[29] Y. Halevi, Control of flexible structures governed by the wave equation using infinite dimensional transfer functions, Journal of Dynamic Systems, Measurement, and Control. 127 (2005) 579–588. https://doi.org/10.1115/1.2098895.

[30] R. Gabai, I. Bucher, Spatial and temporal excitation to generate traveling waves in structures, Journal of Applied Mechanics, Transactions ASME. 77 (2010) 1–11. https://doi.org/10.1115/1.3176999.

[31] L. Sirota, Y. Halevi, Fractional order control of the two-dimensional wave equation, Automatica. 59 (2015) 152–163. https://doi.org/10.1016/j.automatica.2015.06.016.

[32] P.A. Musgrave, Patrick F and Malladi, VVN Sriram and Tarazaga, Generation of traveling waves in a 2D plate for future drag reduction manipulation, in: Special Topics in Structural Dynamics, Volume 6, Springer, 2016: pp. 129-138.

[33] M. Åbom, Modal decomposition in ducts based on transfer function measurements between microphone pairs, Journal of Sound and Vibration. 135 (1989) 95–114. https://doi.org/10.1016/0022-460X(89)90757-8.

[34] S. Allam, M. Åbom, Investigation of damping and radiation using full plane wave decomposition in ducts Journal of Sound and Vibration. 292 (2006) 519–534. https://doi.org/10.1016/j.jsv.2005.08.016.

[35] S. Sack, M. Åbom, On Acoustic Multi-Port Characterisation Including Higher Order Modes, Acta Acust. United with Acust. 102 (2016) 834–850. https://doi.org/10.3813/AAA.918998.

[36] Y. Vered, R. Gabai, I. Bucher, Dispersion based reduced-order model identification and boundary impedance control in a weakly coupled impedance tube, Proceedings of Meetings on Acoustics. 39 (2019) 045005. https://doi.org/10.1121/2.0001215.

[37] M. Bouchard, S. Quednau, Multichannel recursive-least-square algorithms and fast-transversal-filter algorithms for active noise control and sound reproduction systems, IEEE Transactions on Speech and Audio Processing. 8 (2000) 606–618. https://doi.org/10.1109/89.861382.

[38] M. Feldman, Hilbert transform in vibration analysis, Mechanical Systems and Signal Processing. 25 (2011) 735–802. https://doi.org/10.1016/j.ymssp.2010.07.018.

[39] L. Brillouin, Wave Propagation in Periodic Structures: Electric Filters and Crystal Lattices, Dover Publications, 1953.

[40] D. Duhamel, B.R. Mace, M.J. Brennan, Finite element analysis of the vibrations of waveguides and periodic structures, Journal of Sound and Vibration. 294 (2006) 205–220. https://doi.org/10.1016/j.jsv.2005.11.014.

[41] M.H. Hayes, Statistical digital signal processing and modeling, John Wiley & Sons, 2009.

[42] P.C. Hansen, V. Pereyra, and G. Scherer, Least squares data fitting with applications, Johns Hopkins University Press, Baltimore, 2012. https://doi.org/10.1353/book.21076.

[43] Å. Björck, Component-wise perturbation analysis and error bounds for linear least squares solutions, BIT 31 (1991) 237–244. https://doi.org/10.1007/BF01931284.

[44] S.J. Elliot, Signal Processing for Active Control, Elsevier, 2001. https://doi.org/10.1016/b978-0-12-237085-4.x5000-5.

[45] P.C. Hansen, Analysis of discrete ill-posed problems by means of the L-curve, SIAM Review. 34 (1992) 561–580. https://doi.org/10.1137/1034115.

[46] I. Bucher, S.G. Braun, Left eigenvectors: extraction from measurements and physical interpretation, Journal of Applied Mechanics. 64 (1997) 97–105. https://doi.org/10.1115/1.2787300.

[47] J. Achenbach, Wave propagation in elastic solids, North-holland Publishing Company/ American Elsevier, 1973. http://refhub.elsevier.com/S0888-3270(18)30800-8/h0005

[48] J.J. Jensen, On the shear coefficient in Timoshenko's beam theory, Journal of Sound and Vibration. 87 (1983) 621–635. https://doi.org/10.1016/0022-460X(83)90511-4.

[49] A.D. Pierce, R.T Beyer, Acoustics: an introduction to its physical principles and applications, 1989.

[50] A.F. Seybert, B. Soenarko, Error analysis of spectral estimates with application to the measurement of acoustic parameters using random sound fields in ducts, The Journal of the Acoustical Society of America. 69 (1981) 1190–1199. https://doi.org/10.1121/1.385700.

[51] H. Bodén, M. Åbom, Influence of errors on the two-microphone method for measuring acoustic properties in ducts, Journal of the Acoustical Society of America. 79 (1986) 541–549. https://doi.org/10.1121/1.393542.

[52] C. Peng, D. Morrey, P. Sanders, The Measurement of Low Frequency Impedance Using an Impedance Tube, Journal of Low Frequency Noise, Vibration and Active Control. 17 (1998) 1–10. https://doi.org/10.1177/026309239801700101.

[53] S. Braun, Extraction of Periodic Waveforms By Time Domain Averaging, Acustica. 32 (1975) 69–77.